\documentclass[prb,amsmath,twocolumn,showpacs]{revtex4}
\usepackage{bm}
\usepackage{amssymb}
\usepackage{graphicx}
\def\bS{{\mbox{\boldmath $S$}}}
\def\btau{{\mbox{\boldmath $\tau$}}}
\def\rqb{{\rm{qb}}}
\def\rset{{\rm{SET}}}
\def\rint{{\rm{int}}}
\def\b0{{\bf{0}}}
\begin{document}
\title{Quantum mechanical approach to decoherence and relaxation
generated by fluctuating environment}

\author{S.A. Gurvitz}
\email{shmuel.gurvitz@weizmann.ac.il}

\affiliation{Department of Particle Physics,  Weizmann Institute of
Science, Rehovot 76100, Israel and Theoretical Division and CNLS,
Los Alamos National Laboratory, Los Alamos, NM 87545, USA}

\author{D. Mozyrsky}

\affiliation{Theoretical Division, Los Alamos National Laboratory,
Los Alamos, NM 87545, USA}

\date{\today}


\begin{abstract}
We consider an electrostatic qubit, interacting with fluctuating
charge of a single electron transistor (SET) in the framework of an
exactly solvable model. The SET plays role of an environment
affecting the qubits' parameters in a controllable way. We derive the
rate equations describing the dynamics of the entire system for an
arbitrary qubit-SET coupling. Solving these equations we obtain
decoherence and relaxation rates of the qubit, as well as the
spectral density of qubit parameters' fluctuations. We found that in
a weak coupling regime decoherence and relaxation rates are directly
related to the spectral density taken at either zero or Rabi
frequency, depending on which qubit parameter  is fluctuating. In the
latter case our result coincides with that of the spin-boson model in
the  weak coupling limit, despite different origin of the
fluctuations. We show that this relation holds also in the presence
of weak back-action of the qubit on the environment. In case of
strong back-action such a simple relationship no longer holds, even
if qubit-SET coupling is small. It does not hold also in the strong
coupling regime, even in the absence of the back-action. In addition,
we found that our model predicts localization of the qubit in the
strong-coupling regime, resembling that in the spin-boson model.
\end{abstract}

\maketitle

\section{Introduction}
The influence of environment on a single quantum system is the issue
of crucial importance in quantum information science. It is mainly
associated with decoherence, or dephasing, which transforms any pure
state of a quantum system into a statistical mixture. Despite a
large body of theoretical work devoted to decoherence, its mechanism
has not been clarified enough. For instance, how decoherence is
related to environmental noise, in particular in the presence of
back-action of the system on the environment (quantum
measurements). Moreover, decoherence is often intermixed with
relaxation. Although each of them represents an irreversible
process, decoherence and relaxation affect quantum systems in quite
different ways.

In order to establish a relation between the fluctuation spectrum
and decoherence and relaxation rates one needs a model that
describes the effects of decoherence and relaxation in a consistent
quantum mechanical way. An obvious candidate is the spin-boson
model\cite{legg, weiss} which represents the environment as a bath
of harmonic oscillators at equilibrium, where the fluctuations obey
Gaussian statistics \cite{shnir}. Despite its apparent simplicity,
the spin-boson model cannot be solved exactly \cite{weiss}. Also, it
is hard to manipulate the fluctuation spectrum in the framework of
this model. In addition, mesoscopic structures may couple only to a
few isolated fluctuators, like spins, local currents, background
charge fluctuations, etc. This would require models of the
environment, different from the spin-boson model (see for instace
\cite{gass,paladino,tokura,nori2,grishin,machlin,galp,nori1,hart}).
In general, the environment can be out of equilibrium, like a
steady-state fluctuating current, interacting with the qubit
\cite{g2,but,clerk,gb}. This for instance, takes place in the
continuous measurement (monitoring) of quantum systems\cite{kak} and
in the ``control dephasing'' experiments\cite{agu,kouw,neder1}. All
these types on non-Gaussian and non-equilibrium environments
attracted recently a great deal of attention \cite{izhar}.
\begin{figure}
\includegraphics[width=7cm]{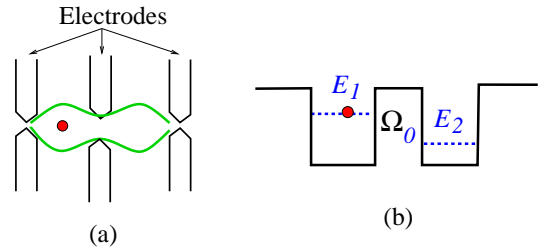}
\caption{Electrostatic qubit, realized by an electron trapped in a
coupled-dot system (a), and its schematic representation by a
double-well (b). $\Omega_0$ denotes the coupling between the two
dots.} \label{fig1}
\end{figure}

In this paper we consider an electrostatic qubit, which can be
viewed as a generic example of two-state systems. It is realized by
an electron trapped in coupled quantum dots \cite{exp2,exp1,exp3},
Fig.~1. Here $E_1$ and $E_2$ denote energies of the electron states
in each of the dots and $\Omega_0$ is a coupling between these
states. It is reasonable to assume that the decoherence of a qubit
is associated with fluctuations of the qubit parameters, $E_{1,2}$
and $\Omega_0$, generated by the environment. Indeed, a stochastic
averaging of the Schr\"odinger equation over these fluctuations
parameters results in the qubit's decoherence, which transfers any
qubit state into a statistical mixture \cite{shao,wang}. In general,
one can expect that the fluctuating environment should result in the
qubit's relaxation, as well, as for instance in the phenomenological
Redfield's description of relaxation in the magnetic resonance
\cite{slich}.

As a quantum mechanical model of the environment we consider a
Single Electron Transistor (SET) capacitively coupled to the qubit,
e.g., Fig.~2. Such setup has been contemplated in numerous solid
state quantum computing architectures where SET plays role of a
readout device \cite{kak,makh,goan,gb} and contains most of the generic
features of a fluctuating non-equilibrium environment. The
discreteness of the electron charge creates fluctuations in the
electrostatic field near the SET. If the electrostatic qubit is
placed near the SET, this fluctuating field should affect the qubit
behavior as shown in Fig.~2. It can produce fluctuations of the
tunneling coupling between the dots (off-diagonal coupling) by
narrowing the electrostatic opening connecting these dots, as in
Fig.~2a, or make the energy levels of the dots fluctuate, as shown
schematically in Fig.~2b. Note that while in some regimes the SET
operates as a measuring device \cite{kak,gb}, in other regimes it
corresponds purely to a source of noise. Indeed, if the energy level
$E_0$, Fig.~2, is deeply inside the voltage bias -- the case we
consider in the beginning,  the SET current is not modulated by the
qubit electron. In this case the SET represents only the fluctuating
environment affecting the qubit behavior (``pure environment''
\cite{kor1}).

\begin{figure}
\includegraphics[width=8cm]{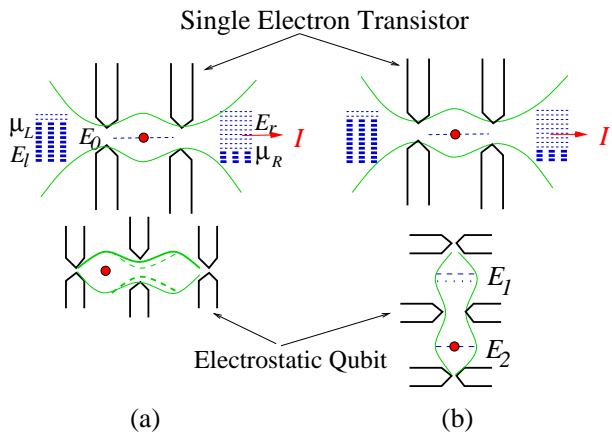}
\caption{Qubit near Single Electron Transistor. Here $E_{l,r}$ and
$E_0$ denote the energy levels in the left (right) reservoirs and in
the quantum dot, respectively, and $\mu_{L,R}$ are the corresponding
chemical potentials. The electric current $I$ generates fluctuations
of the electrostatic opening between two dots (a), or it fluctuates
the energy level of the nearest dot (b).} \label{fig2}
\end{figure}

A similar model of the fluctuating environment has been studied
mostly for small bias (linear response) or for the environment in an
equilibrium. Here, however, we consider  strongly non-equilibrium
case where the bias voltage applied on the SET ($V=\mu_L-\mu_R$) is
much larger than the levels widths and the coupling between the SET
and the qubit. In this limit our model can be solved exactly for
both weak and strong coupling (but is still smaller than the bias
voltage). This constitutes an essential advantage with regard to
perturbative treatments of similar models. For instance, the results
of our model can be compared in different regimes with
phenomenological descriptions used in the literature. Such a
comparison would allow us to determine the regions where these
phenomenological models are valid.

Since our model is very simple in treatment, the decoherence and
relaxation rates can be extracted from the exact solution
analytically, as well as the time-correlator of the electric charge
inside the SET. This would make it possible to establish a relation
between the frequency-dependent fluctuation spectrum of the
environment and the decoherence and the relaxation rates of the
qubit, and to determine how far this relation can be extended. We
expect that such a relation should not depend on a source of
fluctuations. This point can be verified by a comparison with a
similar results obtained for equilibrium environment in the
framework of the spin-boson model \cite{legg,weiss}.

It is also important to understand how the decoherence and
relaxation rates depend on the frequency of the environmental
fluctuations. This problem has been investigated in many
phenomenological approaches for ``classical'' environments at
equilibrium. Yet, there still exists an ambiguity in the literature
related to this point for non-equilibrium environment. For instance,
it was found by Levinson that the decoherence rate, generated by
fluctuations of the energy level in a single quantum dot is
proportional to the spectral density of fluctuations at zero
frequency \cite{levins}. The same result, but for a double-dot
system has been obtained by Rabenstein {\em et al.} \cite{raben}. On
the other hand, it follows from the Redfied's approach that the
corresponding decoherence rate is proportional to the spectral
density at the frequency of the qubit's oscillations (the Rabi
frequency)\cite{slich}. Since our model is the exactly solvable one,
we can resolve this ambiguity and establish the appropriate physical
conditions that can result in different relations of decoherence
rate to the environmental fluctuations.

The most important results of our study are related to the situation
when back-action of the qubit on the environment takes place. This
problem did not receive such a considerable amount of attention in
the literature as, for example, the case of ``inert'' environment.
This is in spite of a fact that the back-action always takes place
in the presence of measurement. There are many questions related to
the effects of a back-action. For instance, what would be a relation
between decoherence (relaxation) of the qubit and the noise spectrum
of the environment? Or, how decoherence is affected by a strong
response of environment? We believe that our model appears to be
more suitable for studying these and other problems related to the
back-action than most of the other existing approaches.

The plan of this paper is as follows: Sect. II presents a
phenomenological description of decoherence and relaxation in the
framework of Bloch equations, applied to the electrostatic qubit.
Sect. III contains description of the model and the quantum
rate-equation formalism, used for its solution. Detailed
quantum-mechanical derivation of these equations for a specific
example is presented in Appendix A. Sect. IV deals with a
configuration where the SET can generate only decoherence of the
qubit. We consider separately the situations when SET produces
fluctuations of the tunneling coupling (Rabi frequency) or of the
energy levels. The results are compared with the SET fluctuation
spectrum, evaluated in Appendix B. Sect. V deals with a
configuration where the SET generates both decoherence and
relaxation of the qubit. Sect. VI is summary.

\section{Decoherence and relaxation of a qubit}

In this section we describe in a general phenomenological framework the
effect of decoherence and relaxation on the qubit behavior. Although
the results are known, there still exists some confusion in the
literature in this issue. We therefore need to define precisely
these quantities and demonstrate how the corresponding decoherenece
and relaxation rates can be extracted from the qubit density matrix.

Let us consider an electrostatic qubit, realized by an electron
trapped in coupled quantum dots, Fig.~1. This system is described by
the following tunneling Hamiltonian
\begin{align}H_\rqb = E_1 a_1^{\dagger}a_{1}+E_2
a_2^{\dagger}a_{2} -\Omega_0 (a_2^{\dagger}a_{1}+
a_1^{\dagger}a_{2}) \label{a0}
\end{align}
where $a^\dagger_{1,2}, a_{1,2}$ are the creation and annihilation
operators of the electron in the first or in the second dot. For
simplicity we consider electrons as spinless fermions. In addition,
we assume that $a_1^{\dagger}a_{1}+a_2^{\dagger}a_{2}=1$, so that
only one electron is present in the double-dot. The electron wave
function can be written as \begin{align}|\Psi (t)\rangle = \left
[b^{(1)}(t)a_1^{\dagger} + b^{(2)}(t)a_2^{\dagger}\right ]
|\b0\rangle\label{a01}
\end{align}
where $b^{(1,2)}(t)$ are the probability amplitudes for finding the
electron in the first or second well, obtained from the
Schr\"odinger equation $i\partial_t |\Psi (t)\rangle =H_\rqb |\Psi
(t)\rangle$ (we adopt the units where $\hbar =1$ and the electron
charge $e=1$). The corresponding density matrix,
$\sigma_{jj'}(t)=b^{(j)}(t)b^{{(j')}*}(t)$, with $j,j'=\{1,2\}$, is
obtained from the equation $i\partial_t\,\sigma =[H,\sigma ]$. This
can be written explicitly as
\begin{subequations}
\label{a02}
\begin{eqnarray}
\label{a02a}
\dot{\sigma}_{11}&=&i\Omega_0(\sigma_{21}-\sigma_{12})\\
\label{a02b}
\dot{\sigma}_{12}&=&-i\epsilon\sigma_{12}+i\Omega_0(1-2\sigma_{11})\,
,
\end{eqnarray}
\end{subequations}
where $\sigma_{22}(t)=1-\sigma_{11}(t)$,
$\sigma_{21}(t)=\sigma_{12}^*(t)$ and $\epsilon =E_1-E_2$. Solving
these equations one easily finds that the electron oscillates
between the two dots (Rabi oscillations) with frequency
$\omega_R=\sqrt{4\Omega_0^2+\epsilon^2}$. For instance, for the
initial conditions $\sigma_{11}(0)=1$ and $\sigma_{12}(0)=1$, the
probability of finding the electron in the second dot is
$\sigma_{22}(t)=2(\Omega_0/\omega_R)^2(1-\cos\omega_R t)$. This
result shows that for $\epsilon\gg\Omega_0$ the amplitude of the
Rabi oscillations is small, so the electron remains localized in its
initial state.

The situation is different when the qubit interacts with the
environment. In this case the (reduced) density matrix of the qubit
$\sigma (t)$ is obtained by tracing out the environment variables
from the total density matrix. The question is how to modify
Eqs.~(\ref{a02}), written for an isolated qubit, in order to obtain
the reduced density matrix of the qubit, $\sigma (t)$. In
general one expects that the environment could affect the qubit in
two different ways. First, it can destroy the off-diagonal elements
of the qubit density matrix. This process is usually referred to as
decoherence (or dephasing). It can be accounted for
phenomenologically by introducing an additional (damping) term in
Eq.~(\ref{a02b}),
\begin{align}
\dot{\sigma}_{12}=-i\epsilon\sigma_{12}+i\Omega_0(1-2\sigma_{11})
-{\Gamma_d\over 2}\sigma_{12}\, \label{a03}
\end{align}
where $\Gamma_d$ is the decoherence rate. As a result the qubit
density-matrix $\sigma (t)$ becomes a statistical mixture in the
stationary limit,
\begin{align}
\sigma(t) \stackrel{\small t\to\infty}{\longrightarrow}
\left (\begin{array}{cc}1/2&0\\
0&1/2\end{array}\right )\, . \label{a04}
\end{align}
This happens for any initial conditions and even for large level
displacement, $\epsilon\gg\Omega_0,\Gamma_d$ (provided that
$\Omega_0\not =0$). Note that the statistical mixture (\ref{a04}) is
proportional to the unity matrix and therefore it remains the same
in any basis.

Secondly, the environment can put the qubit in its ground state, for
instance via photon or phonon emission. This process is usually
referred to as relaxation. For a symmetric qubit we would have
\begin{align}
\sigma(t) \stackrel{\small t\to\infty}{\longrightarrow}\left (
\begin{array}{cc}1/2&1/2\\
1/2&1/2\end{array} \right )\, . \label{a05}
\end{align}
In contrast with decoherence, Eq.~(\ref{a04}), the relaxation
process puts the qubit into a pure state. That implies that the
corresponding
density matrix can be always written as $\delta_{1i}\delta_{1j}$
in a certain basis (the basis of the qubit eigenstates). This is
in fact the essential difference between decoherence and relaxation.
With respect to elimination of the off-diagonal density matrix elements,
note that  relaxation would eliminate these terms only in the qubit's
eigenstates basis. In contrast, decoherence eliminates the off-diagonal
density matrix element in any basis  (Eq.~(\ref{a04})).

In fact, if the environment has some energy, it can put the qubit
into an exited state. However, if the qubit is finally in a pure
state, such excitation process generated by the environment affects
the qubit in the same way as relaxation: it eliminates the
off-diagonal density matrix elements only in a certain qubit's
basis. Therefore excitation of the qubit can be described
phenomenologically on the same footing as relaxation.

It is often claimed that decoherence is associated with an absence
of energy transfer between the system and the environment, in
contrast with relaxation (excitation). This distinction is not
generally valid. For instance, if the initial qubit state
corresponds to the electron in the state $|E_2\rangle$, Fig.~1, the
final state after decoherence corresponds to an equal distribution
between the two dots, $\langle E\rangle =(E_1+E_2)/2$.  In the case
of $E_1\gg E_2$, this process would require a large energy transfer
between the qubit and the environment. Therefore decoherence can be
consistently defines as a process leading to a statistical mixture,
where all states of the system have equal probabilities (as in
Eq.~(\ref{a04})).

The relaxation (excitation) process can be described most simply by
diagonalizing the qubit Hamiltonian, Eqs.~(\ref{a0}), to obtain
$H_\rqb = E_+ a_+^{\dagger}a_{+}+E_- a_-^{\dagger}a_{-}$, where the
operators $a_{\pm}$ are obtained by the corresponding rotation of
the operators $a_{1,2}$ \cite{kor1}. Here $E_+$ and $E_-$ are the
ground (symmetric) and excited (antisymmetric) state energies. Then
the relaxation process can be described phenomenologically in the
new qubit basis $|\pm\rangle= a_{\pm}^\dagger|\b0 \rangle$ as
\begin{subequations}
\label{a06}
\begin{eqnarray}
  && \dot\sigma_{--}(t)=-\Gamma_r\sigma_{--}(t)
  \label{a06a}\\
  && \dot\sigma_{+-}(t)=i(E_--E_+)\sigma_{+-}(t)
  -{\Gamma_r\over 2}\sigma_{+-}(t)\, ,
  \label{a06b}
 \end{eqnarray}
\end{subequations}
where $\sigma_{++}(t)=1-\sigma_{--}(t)$,
$\sigma_{-+}(t)=\sigma^*_{+-}(t)$ and $\Gamma_r$ is the relaxation
rate.

In order to add decoherence, we return to the original qubit basis
$|1,2\rangle= a_{1,2}^\dagger|\b0\rangle$ and add the damping term
to the equation for the off-diagonal matrix elements,
Eq.~(\ref{a03}). We arrive at the quantum rate equation describing
the qubit's behavior in the presence of both decoherence and
relaxation \cite{kor1,gfmb},
\begin{widetext}
\begin{subequations}
\label{a07}
\begin{eqnarray}\label{a07a}
&&\dot\sigma_{11}=i\Omega_0(\sigma_{21}-\sigma_{12}) -\Gamma_r\,
{\kappa\epsilon \over2 \tilde\epsilon}\, (\sigma_{12}+\sigma_{21})
-{\Gamma_r\over4}\left[1+\left({\epsilon\over\tilde\epsilon
}\right)^2\right](2\sigma_{11}-1) +\Gamma_r{\epsilon \over
2\tilde\epsilon}
 \\[5pt]
  &&\dot\sigma_{12}=-i\epsilon\sigma_{12}
  +\left[i\Omega_0+\Gamma_r{\kappa\epsilon \over 2\tilde\epsilon}\right]
  (1-2\sigma_{11})
+\Gamma_r\left[\kappa -{1\over2}\sigma_{12}
  -\kappa^2(\sigma_{12}+\sigma_{21})\right]
-{\Gamma_d\over2}\sigma_{12}\, , \label{a07b}
\end{eqnarray}
\end{subequations}
\end{widetext} where $\tilde\epsilon =(\epsilon^2+4\Omega_0^2)^{1/2}$
and $\kappa =\Omega_0/\tilde\epsilon$. In fact, these equations can
be derived in the framework of a particular model, representing an
electrostatic qubit interacting with the point-contact detector and
the environment, described by the Lee model Hamiltonian \cite{gfmb}.

Equations~(\ref{a07}) can be rewritten in a simpler form by mapping
the qubit density matrix $\sigma
=\{\sigma_{11},\sigma_{12},\sigma_{21}\}$ to a ``polarization''
vector $\bS(t)$ via $\sigma (t)=[1+\btau\cdot \bS (t)]/2$, where
$\tau_{x,y,z}$ are the Pauli matrices. For instance, one obtains for
the symmetric case, $\epsilon =0$,
\begin{subequations}
\label{a08}
\begin{eqnarray}
  &&\dot S_z=-{\Gamma_r\over2}S_z-2\Omega_0\, S_y
   \label{a08a}\\
   &&\dot S_y=2\Omega_0\, S_z-{\Gamma_d+\Gamma_r\over2}S_y
 \label{a9b}\\
 &&\dot S_x=-{\Gamma_d+2\Gamma_r\over2}(S_x-\bar S_x)
  \label{a08c}
\end{eqnarray}
\end{subequations}
where $\bar S_x=S_x(t\to\infty )=2\Gamma_r/(\Gamma_d+2\Gamma_r)$.
One finds that Eqs.~(\ref{a08}) have a form of the Bloch equations
for spin-precession in the magnetic field \cite{slich}, where the
effect of environment is accounted for by two relaxation times for
the different spin components: the longitudinal $T_1$ and the
transverse $T_2$, related to $\Gamma_d$ and $2\Gamma_r$ as
\begin{align}
T_1^{-1}={\Gamma_d+2\Gamma_r\over 2},~~~ {\mbox{and}}~~~
T_2^{-1}={\Gamma_d+\Gamma_r\over 2}\, , \label{aa088}
\end{align}
The corresponding damping rates, the so-called ``depolarization''
($\Gamma_{1}=1/T_{1}$) and the ``dephasing'' ($\Gamma_{2}=1/T_{2}$)
are used for phenomenological description of two-level systems
\cite{ithier}. However, neither $\Gamma_1$ nor $\Gamma_2$ taken
alone would drive the qubit density matrix into a statistical
mixture Eq.~(\ref{a04}) or into a pure state Eq.~(\ref{a05}).

In contrast, our definition of decoherence and relaxation
(excitation) is associated with two opposite effects of the
environment on the qubit: the first drives it into a statistical
mixture, whereas the second drives it into a pure state. We expect
therefore that such a natural distinction between decoherence and
relaxation would be more useful for finding a relation between these
quantities and the environmental behavior than other alternative
definitions of these quantities existing in the literature.

In general, the two rates, $\Gamma_{d,r}$, introduced in
phenomenological equations~(\ref{a07}), (\ref{a08}), are consistent
with our definitions of decoherence and relaxation. The only
exception is the case of $\Gamma_r=0$ and $\Omega_0=0$, where are no
transitions between the qubit's states even in the presence of the
environment (``static'' qubit). One easily finds from
Eqs.~(\ref{a02a}), (\ref{a03}) that $\sigma_{12}(t)\to 0$ for
$t\to\infty$, whereas the diagonal density-matrix elements of the
qubit remain unchanged (so-called ``pure dephasing''
\cite{paladino,ithier}):
\begin{align}
\sigma(t) \stackrel{\small t\to\infty}{\longrightarrow}
\left (\begin{array}{cc}\sigma_{11}(0)&0\\
0&\sigma_{22}(0)\end{array}\right )\, . \label{static}
\end{align}
Thus, if the initial probabilities of finding the qubit in each of
its states are not equal, $\sigma_{11}(0)\not =\sigma_{22}(0)$, then
the final qubit state is neither a mixture nor a pure state, but a
combination of the both. It implies that $\Gamma_d$ in
Eqs.~(\ref{a07}) would also generate relaxation (excitation) of the
qubit. Note that in this case the off-diagonal density-matrix
elements, absent in Eq.(\ref{static}), would reappear in a different
basis. This implies that the ``pure dephasing''
\cite{paladino,ithier} occurs only in a particular basis.

Let us evaluate the probability of finding the electron in the first
dot, $\sigma_{11}(t)$. Solving Eqs.~(\ref{a08}) for the initial
conditions $\sigma_{11}(0)=1$, $\sigma_{12}(0)=0$, we find
\cite{gfmb}:
\begin{align}
\sigma_{11}(t)=\frac{1}{2} +{e^{-\Gamma_rt/2}\over 4}\left (C_1
e^{-e_-t} +C_2e^{-e_+t}\right ) \label{aa08}
\end{align}
where $e_{\pm}={1\over4}(\Gamma_d\pm\tilde\Omega)$, $\tilde\Omega
=\sqrt{\Gamma_d^2-64\Omega_0^2}$ and
$C_{1,2}=1\pm(\Gamma_d/\tilde\Omega)$. Solving the same equations in
the limit of $t\to\infty$, we find that the steady-state qubit
density matrix is
\begin{align}
\sigma(t) \stackrel{\small t\to\infty}{\longrightarrow}\left
(\begin{array}{cc}
1/2&\Gamma_r/(\Gamma_d+2\Gamma_r)\cr\Gamma_r/(\Gamma_d+2\Gamma_r)&1/2
\end{array}\right )\, .\label{a09}
\end{align}
Thus the off-diagonal elements of the density matrix can provide us
with a ratio of relaxation to decoherence rates\cite{gfmb}.

\section{Description of the model}
Consider the setup shown in Fig.~2. The entire system can be
described by the following tunneling Hamiltonian, represented by a
sum of the qubit and SET Hamiltonians and the interaction term,
$H=H_\rqb +H_\rset +H_\rint $. Here $H_\rqb$ is given by
Eq.~(\ref{a0}) and describes the qubit. The second term, $H_\rset$,
describes the single-electron transistor. It can be written as
\begin{align}
H_\rset&=\sum_lE_lc_l^\dagger c_l+\sum_rE_rc_r^\dagger
c_r+E_0c_0^\dagger c_0\nonumber\\&+\sum_{l,r}(\Omega_lc^\dagger_l
c_0 +\Omega_rc^\dagger_r c_0+H.c.)\, , \label{a1a}
\end{align}
where $c_{l,r}^\dagger$ and $c_{l,r}$ are the creation and
annihilation electron operators in the state $E_{l,r}$ of the right
or left reservoir; $c_{0}^\dagger$ and $c_{0}$ are those for the
level $E_0$ inside the quantum dot; and $\Omega_{l,r}$ are the
couplings between the level $E_0$ and the level $E_{l,r}$ in the
left (right) reservoir. In order to avoid too lengthy formulaes, our
summation indices $l,r$ indicate simultaneously the left and the
right leads of the SET, where the corresponding summation is carried
out. As follows from the Hamiltonian~(\ref{a1a}), the quantum dot of
the SET contains only one level ($E_0$). This assumption has been
implied only for the sake of simplicity for our presentation, although
our approach is well suited for a case of $n$ levels inside the SET,
$E_0c_0^\dagger c_0\to\sum_nE_nc_n^\dagger c_n$, and even when
the interaction between these levels is included (providing that the
latter is much less or much larger than the bias $V$) \cite{ieee,gmb}.
We also assumed a weak energy
dependence of the couplings $\Omega_{l,r}\simeq\Omega_{L,R}$.

The interaction between the qubit and the SET,  $H_\rint$, depends
on a position of the SET with respect to the qubit. If the SET is
placed near the middle of the qubit, Fig.~2a, then the tunneling
coupling between two dots of the qubit in Eq.~(\ref{a0}) decreases,
$\Omega_0\to\Omega_0-\delta\Omega_0$, whenever the quantum dot of
the SET is occupied by an electron. This is due to the electron's
repulsive field. In this case the interaction term can be written as
\begin{align}
H_\rint=\delta\Omega\, c_0^\dagger c_0(a_1^\dagger a_2+a_2^\dagger
a_1)\, . \label{a1b}
\end{align}
On the other hand, in the configuration shown in Fig.~2b where the
SET is placed near one of the dots of the qubit, the electron
repulsive field displaces the qubit energy levels by $\Delta E=U$.
The interaction terms in this case can be written as
\begin{align}
H_\rint =U\, a_1^\dagger a_1 c_0^\dagger c_0 \, .\label{a1c}
\end{align}

Consider the initial state where all the levels in the left and the
right reservoirs are filled with electrons up to the Fermi levels
$\mu_{L,R}$ respectively. This state will be called the ``vacuum''
state $|\b0 \rangle$. The wave function for the entire system can be
written as
\begin{widetext}
\begin{eqnarray}
&&|\Psi (t)\rangle = \left [ b^{(1)}(t)a_1^\dagger + \sum_l
b^{(1)}_{0l}(t)a_1^\dagger
           c_{0}^{\dagger}c_{l}+ \sum_{l,r} b^{(1)}_{rl}(t)
           a_1^\dagger c_{r}^{\dagger}c_{l}
+\sum_{l<l',r} b^{(1)}_{0rll'}(t)a_1^\dagger
           c_{0}^{\dagger}c_{r}^{\dagger}c_{l}c_{l'}+\cdots\right.
           \nonumber\\
&&\left.+b^{(2)}(t)a_2^\dagger + \sum_l b^{(2)}_{0l}(t)a_2^\dagger
           c_{0}^{\dagger}c_{l}+ \sum_{l,r} b^{(2)}_{rl}(t)
           a_2^\dagger c_{r}^{\dagger}c_{l}
+\sum_{l<l',r} b^{(2)}_{0rll'}(t)a_2^\dagger
           c_{0}^{\dagger}c_{r}^{\dagger}c_{l}c_{l'}
+\ldots \right ] |\b0\rangle, \label{a2}
\end{eqnarray}
\end{widetext}
where $b^{(j)}(t)$, $b^{(j)}_\alpha(t)$ are the probability
amplitudes to find the entire system in the state described by the
corresponding creation and annihilation operators. These amplitudes
are obtained from the Schr\"odinger equation $i|\dot\Psi(t)\rangle=
H|\Psi(t)\rangle$, supplemented with the initial condition
$b^{(1)}(0)=p_1$, $b^{(2)}(0)=p_2$, and $b_{\alpha}^{(j)}(0)=0$,
where $p_{1,2}$ are the amplitudes of the initial qubit state.

Note that Eq.~(\ref{a2}) implies a fixed electron number ($N$) in
the reservoirs. At the first sight it would lead to depletion of the
left reservoir of electrons over the time. Yet in the limit of
$N\to\infty$ (infinite reservoirs) the dynamics of an entire system
reaches its steady state before such a depletion takes
place\cite{gp,g1}.

The behavior of the qubit and the SET is given by the reduced
density matrix, $\sigma_{ss'}(t)$. It is obtained from the entire
system's density matrix $|\Psi (t)\rangle\langle\Psi (t)|$ by
tracing out the (continuum) reservoir states. The space of such a
reduced density matrix consists of four discrete states
$s,s'=a,b,c,d$, shown schematically in Fig.~3 for the setup of
Fig.~2a. The corresponding density-matrix elements are directly
related to the amplitudes $b(t)$, for instance,
\begin{widetext}
\begin{subequations}
\label{a3}
\begin{eqnarray}
\sigma_{aa}(t)&=&|b^{(1)}(t)|^2+\sum_{l,r}|b^{(1)}_{lr}(t)|^2
+\sum_{l<l',r<r'}|b^{(1)}_{rr'll'}(t)|^2+\cdots \label{a3a}\\
\sigma_{dd}(t)&=&\sum_l|b_{0l}^{(2)}(t)|^2+\sum_{l<l',r}|b^{(2)}_{0rll'}(t)|^2
+\sum_{l<l'<l'',r<r'}|b^{(2)}_{0rr'll'}(t)|^2+\cdots\label{a3d}\\
\sigma_{bd}(t)&=&\sum_lb_{0l}^{(1)}(t)b_{0l}^{(2)*}(t)
+\sum_{l<l',r}b^{(1)}_{0rll'}(t)b^{(2)*}_{0rll'}(t)
+\sum_{l<l'<l'',r<r'}b^{(1)}_{0rr'll'}(t)b^{(2)*}_{0rr'll'}(t)+\cdots
. \label{a3c}
\end{eqnarray}
\end{subequations}
\end{widetext}
In was shown in\cite{gp,g1} that the trace over the
reservoir states in the system's density matrix can be performed in
the large bias limit (strong non-equilibrium limit)
\begin{align}
V=\mu_L-\mu_R\gg \Gamma, \Omega_0, U
\label{aa3}
\end{align}
where the level (levels) of the SET carrying the current are far away
from the chemical potentials, and $\Gamma$ is the width of the level
$E_0$. In this derivation we assumed
only weak energy dependence of the transition amplitudes
$\Omega_{l,r}\equiv\Omega_{L,R}$ and the density of the reservoir
states, $\rho(E_{l,r})=\rho_{L,R}$. As a result we arrive at
Bloch-type rate equations for the reduced density matrix without any
additional assumptions. The general form of these equations is \cite
{g1,gmb}
\begin{align}
\dot\sigma_{jj'}&=i(E_{j' }-E_j)\sigma_{jj'} + i
\sum_{k}\left(\sigma_{jk}\tilde\Omega_{k\to j'}
-\tilde\Omega_{j\to k} \sigma_{kj'}\right )
\nonumber\\
&-{\sum_{k,k'}}{\cal P}_2 \pi\rho(\sigma_{jk}\Omega_{k\to
k'}\Omega_{k'\to j'} +\sigma_{kj'}\Omega_{k\to k'}\Omega_{k'\to j})
\nonumber\\&+\sum_{k,k'}{\cal P}_2
\pi\rho\,(\Omega_{k\to j}\Omega_{k'\to j'}+ \Omega_{k\to
j'}\Omega_{k'\to j}) \sigma_{kk'} \label{a4}
\end{align}
\noindent Here $\Omega_{k\to k'}$ denotes the
single-electron hopping amplitude that generates the $k\to k'$
transition. We distinguish between the amplitudes $\tilde\Omega$
describing single-electron hopping between isolated states and
$\Omega$ describing transitions between isolated and continuum
states. The latter can generate transitions between the isolated
states of the system, but only indirectly, via two consecutive jumps
of an electron, into and out of the {\em continuum} reservoir states
(with the density of states $\rho$). These transitions are
represented by the third and the fourth terms of Eq.~(\ref{a4}). The
third term describes the transitions ($k\to k' \to j$) or ($k\to k'
\to j'$), which cannot change the number of electrons in the
collector. The fourth term describes the transitions ($k\to j$ and
$k' \to j'$) or ($k\to j'$ and $k' \to j$) which increase the number
of electrons in the collector by one. These two terms of
Eq.~(\ref{a4}) are analogues of the ``loss'' (negative) and the
``gain'' (positive) terms in the classical rate equations,
respectively. The factor ${\cal P}_2=\pm 1$ in front of these terms
is due anti-commutation of the fermions, so that ${\cal P}_2=-1$
whenever the loss or the gain terms in Eq.~(\ref{a4}) proceed
through a two-fermion state of the dot. Otherwise ${\cal P}_2=1$.

Note that the reduction of the time-dependent Schr\"odinger
equation, $i|\dot\Psi(t)\rangle= H|\Psi(t)\rangle$, to
Eqs.~(\ref{a4}) is performed in the limit of large bias without
explicit use of any Markov-type or weak coupling approximations. The
accuracy of these equations is respectively $\max (\Gamma
,\Omega_0,U,T)/|\mu_{L,R}-E_j|$. A detailed example of this
derivation is presented in Appendix A for the case of resonant
tunneling through a single level. The derivation there and in
Refs.\cite{gp,g1} were performed by assuming zero temperature in the
leads, $T=0$. Yet, this assumption is not important in the case of
large bias, providing the levels carrying the current are far away
from the Fermi energies, $|\mu_{L,R}-E_j|\gg T$.

\section{No back-action on the environment}

\subsection{Fluctuation of the tunneling coupling}
Now we apply Eqs.~(\ref{a4}) to investigate the qubit's behavior in
the configurations shown in Fig.~2. First we consider the SET placed
near the middle of the qubit, Figs.~2a,3. In this case the electron
current through the SET will influence the coupling between two dots
of the qubit, making it fluctuate between the values $\Omega_0$ and
$\Omega'_0=\Omega_0-\delta\Omega$. The corresponding rate equations
can be written straightforwardly from Eqs.~(\ref{a4}). One finds,
\begin{subequations}
\label{a5}
\begin{eqnarray}
\dot\sigma_{aa}&=&-\Gamma_L\sigma_{aa}+\Gamma_R\sigma_{bb}
-i\Omega_0(\sigma_{ac}-\sigma_{ca}),\label{a5a}\\
\dot\sigma_{bb}&=&-\Gamma_R\sigma_{bb}+\Gamma_L\sigma_{aa}
-i\Omega'_0(\sigma_{bd}-\sigma_{db}),\label{a5b}\\
\dot\sigma_{cc}&=&-\Gamma_L\sigma_{cc}+\Gamma_R\sigma_{dd}
-i\Omega_0(\sigma_{ca}-\sigma_{ac}),\label{a5c}\\
\dot\sigma_{dd}&=&-\Gamma_R\sigma_{dd}+\Gamma_L\sigma_{cc}
-i\Omega'_0(\sigma_{db}-\sigma_{bd}),\label{a5d}\\
 \dot\sigma_{ac}&=&-i\epsilon_0\sigma_{ac} -i\Omega_0
(\sigma_{aa}-\sigma_{cc}) -\Gamma_L\sigma_{ac}\nonumber\\
&&~~~~~~~~~~~~~~~~~~~~~~~~~~~~~~~~~~+\Gamma_R\sigma_{bd},\label{a5e} \\
\dot\sigma_{bd}&=&-i\epsilon_0\sigma_{bd}-
i\Omega'_0(\sigma_{bb}-\sigma_{dd})-\Gamma_R\sigma_{bd}\nonumber\\
&&~~~~~~~~~~~~~~~~~~~~~~~~~~~~~~~~~~+\Gamma_L\sigma_{ac},
\label{a5f}
\end{eqnarray}
\end{subequations}
where $\Gamma_{L,R}=2\pi |\Omega_{L,R}|^2\rho_{L,R}$ are the
tunneling rates from the reservoirs and $\epsilon_0 =E_1-E_2$.
\begin{figure}
\includegraphics[width=9cm]{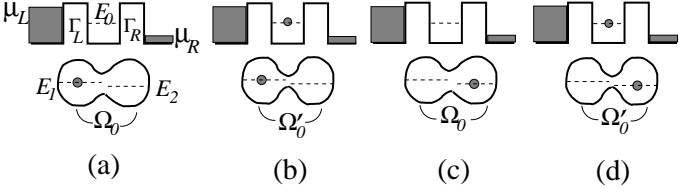}
\caption{The available discrete states of the entire system
corresponding to the setup of Fig.~2a. $\Gamma_{L,R}$ denote the
tunneling rates to the corresponding reservoirs and
$\Omega'_0=\Omega_0-\delta\Omega$.} \label{fig3}
\end{figure}

These equations display explicitly the time evolution of the SET and
the qubit. The evolution of the former is driven by the first two
terms in Eqs.~(\ref{a5a})-(\ref{a5d}). They generate
charge-fluctuations inside the quantum dot of the SET (the
transitions a$\longleftrightarrow$b and c$\longleftrightarrow$d),
described by the ``classical'' Boltzmann-type dynamics. The qubit's
evolution is described by the Bloch-type terms (c.f.
Eqs.~(\ref{a02})), generating the qubit transitions
(a$\longleftrightarrow$c and b$\longleftrightarrow$d). Thus
Eqs.~(\ref{a5}) are quite general, since they described fluctuations
of the tunneling coupling driven by the Boltzmann-type dynamics.

The resulting time evolution of the qubit is given by the qubit
(reduced) density matrix:
\begin{subequations}
\label{aa5}
\begin{eqnarray}
\sigma_{11}(t)&=&\sigma_{aa}(t)+\sigma_{bb}(t)\, ,\label{aa5a}\\
\sigma_{12}(t)&=&\sigma_{ac}(t)+\sigma_{bd}(t)\, ,\label{aa5b}
\end{eqnarray}
\end{subequations}
and $\sigma_{22}(t)=1-\sigma_{11}(t)$.

Similarly, the charge fluctuations of SET are determined by the
probability of finding the SET occupied,
\begin{eqnarray}
P_1(t)&=&\sigma_{bb}(t)+\sigma_{dd}(t)\, .\label{aa6}
\end{eqnarray}
It is given by the equation
\begin{align}
\dot P_1(t)=\Gamma_L-\Gamma P_1(t)\, , \label{a7} \end{align}
obtained straightforwardly from  Eqs.~(\ref{a5}). Here $\Gamma
=\Gamma_L+\Gamma_R$ is the total width. The same equation for
$P_1(t)$ can be obtained if the qubit is decoupled from the SET
($\delta\Omega =0$). Thus there is no back-action of the qubit on
the charge fluctuations inside the SET in the limit of large bias
voltage.

Consider first the stationary limit, $t\to\infty$, where $\dot
P_1(t)\to 0$ and $\dot\sigma (t)\to 0$. It follows from
Eq.~(\ref{a7}) that the probability of finding the SET occupied in
this limit is $\bar P_1=\Gamma_L/\Gamma$. This implies that the
fluctuations of the coupling $\Omega_0$, induced
by the SET, would take place around the average value $\Omega
=\Omega_0-\bar P_1\, \delta\Omega$.

With respect to the qubit in the stationary limit, one easily
obtains from Eqs.~(\ref{a5}) that the qubit density matrix always
becomes the statistical mixture (\ref{a04}), when $t\to\infty$. This
takes place for any initial conditions and any values of the qubit
and the SET parameters. Therefore the effect of the fluctuating
charge inside the SET does not lead to relaxation of the qubit, but
rather to its decoherence.

It is important to note, however, that for the aligned qubit,
$\epsilon =0$, the decoherence due to fluctuations of the tunneling
coupling $\Omega_0$ is not complete. Indeed, it follows from
Eqs.~(\ref{a5}) that $d/dt [{\mbox{Re}}\ \sigma_{12}(t)]=0$. The
reason is that the corresponding operator, $a_1^\dagger
a_2+a_2^\dagger a_1$ commutes with the total Hamiltonian $H=H_\rqb
+H_\rset +H_\rint$, Eqs.~(\ref{a0}), (\ref{a1a}) and (15), for
$E_1=E_2$. As a result, ${\mbox{Re}}\ \sigma_{12}(t)={\mbox{Re}}\
\sigma_{12}(0)$.

In order to determine the decoherence rate analytically, we perform
a Laplace transform on the density matrix, $\tilde\sigma
(E)=\int_0^\infty\sigma (t)\exp (-iEt)dE$. Then solving
Eq.~(\ref{a5}) we can determine the decoherence rate from the
locations of the poles of $\tilde\sigma (E)$ in the complex
$E$-plane. Consider for instance the case of $\epsilon_0 =0$ and the
symmetric SET, $\Gamma_L=\Gamma_R=\Gamma/2$. One finds from
Eqs.~(\ref{a5}) and (\ref{aa5a}) that
\begin{align}
\tilde\sigma_{11}(E)&= \frac{i}{2E}
 +\frac{i(E-2\Omega +i\Gamma)}{\displaystyle 4(E-2\Omega+i\Gamma /2 )^2
 +\Gamma^2-(2\,\delta\Omega)^2}\nonumber\\[5pt]
&+\frac{i(E+2\Omega +i\Gamma)}{\displaystyle 4(E+2\Omega+i\Gamma /2 )^2
+\Gamma^2-(2\,\delta\Omega)^2}
 \, .
\label{a6} \end{align}
Upon performing the inverse
Laplace transform,
\begin{align}
\sigma_{11}(t)=\int\limits_{-\infty +i 0}^{\infty +i
0}\tilde\sigma_{11} (E)\,e^{-iEt}\,{dE\over 2\pi i}, \label{aa7}
\end{align}
and closing the integration contour around the poles of the
integrand, we obtain for $\Gamma > 2\delta\Omega$ and $t\gg
1/\Gamma$
\begin{align}
\sigma_{11}(t)-(1/2)\propto e^{-( \Gamma
-\sqrt{\Gamma^2-4\delta\Omega^2})t/2}\sin (2\Omega\, t). \label{a8}
\end{align}
Comparing this result with Eq.~(\ref{aa08}) we find that the
decoherence rate is
\begin{align}
\Gamma_d=2\left ( \Gamma -\sqrt{\Gamma^2-4\delta\Omega^2}\right)
\stackrel{\small \Gamma\gg\delta\Omega}{\longrightarrow}
(2\delta\Omega)^2/\Gamma\, . \label{a9}
\end{align}
For $\epsilon_0\not =0$ and $\epsilon_0, \Gamma\ll\Omega$ the decoherece
rate $\Gamma_d$ is multiplied by an
additional factor $[1-(\epsilon_0/2\Omega)^2]$.

In a general case, $\Gamma_L\not =\Gamma_R$, we obtain in the same
limit ($\Gamma_{L,R}\gg\delta\Omega$) for the decoherence rate:
\begin{align}
\Gamma_d={(4\,\delta\Omega)^2\over\displaystyle
1+\left({\epsilon_0\over
2\Omega}\right)^2}{\Gamma_L\Gamma_R\over(\Gamma_L+\Gamma_R)^3}
\label{aa9}
\end{align} It is interesting to compare this result with the
fluctuation spectrum of the charge inside the SET, Eq.~(\ref{ac6}),
Appendix B. We find
\begin{align}
\Gamma_d=2\,(\delta\omega_R)^2\,S_Q(0)\, ,\label{a10}
\end{align}
where $\omega_R=\sqrt{4\Omega^2+\epsilon_0^2}$ is the Rabi
frequency. The latter represents the energy splitting in the
diagonalized qubit Hamiltonian. Thus $\delta\omega_R$ corresponds to
the amplitude of energy level fluctuations in a single dot.

Although Eq.~(\ref{a10}) has been obtained for small fluctuations
$\delta\omega_R$, it might be approximately correct even if
$\delta\omega_R$ is of the order of $\Gamma$. It is demonstrated in
Fig.~4, where we compare $\sigma_{11}(t)$ and $\sigma_{12}(t)$,
obtained from Eqs.~(\ref{a5}) and (\ref{aa5}) (solid line) with
those from Eqs.~(\ref{a02a}) and (\ref{a03}) (dashed line) for the
decoherence rate $\Gamma_d$ given by Eq.~(\ref{a10}). The initial
conditions correspond to $\sigma_{11}(0)=1$ and $\sigma_{12}(0)=0$
(respectively, $\sigma_{aa}(0)=\Gamma_R/\Gamma$ and
$\sigma_{bb}(0)=\Gamma_R/\Gamma$).

In the case of aligned qubit, however, ${\mbox{Re}}\
\sigma_{12}(t)={\mbox{Re}}\ \sigma_{12}(0)$, as was explained above.
On the other hand, one always obtains from (\ref{a02a}) and
(\ref{a03}) that ${\mbox{Re}}\, [\sigma_{12}(t\to\infty )]=0$.
Therefore the phenomenological Bloch equations are not applicable
for evaluation of ${\mbox{Re}}\, [\sigma_{12}(t)]$, even in the weak
coupling limit (besides the case of ${\mbox{Re}}\,
[\sigma_{12}(t=0)] =0$).

In the large coupling regime ($\delta\Omega\gg\Gamma$) the
phenomenological Bloch equations, Eqs.~(\ref{a02a}) and (\ref{a03}),
cannot be used, as well. Consider for simplicity the case of
$\epsilon =0$ and $\Gamma_{L,R}=\Gamma /2$. Then one finds from
Eq.~(27) that the damping oscillations between the two dots take
place at two different frequencies,
$2\Omega\pm\sqrt{(\delta\Omega)^2-(\Gamma/ 2)^2}$, instead of the
one frequency, $\omega_R=2\Omega$, given the Bloch equations.
Moreover, Eq.~(\ref{a10}) does not reproduce the decoherence
(damping) rate in this limit. Indeed, one obtains from Eq.~(28) that
the decoherence rate $\Gamma_d =2\Gamma$ for $\delta\Omega
>\Gamma /2$, so $\Gamma_d$  does not depend on the coupling ($\delta
\Omega$) at all.
\begin{figure}
\includegraphics[width=9cm]{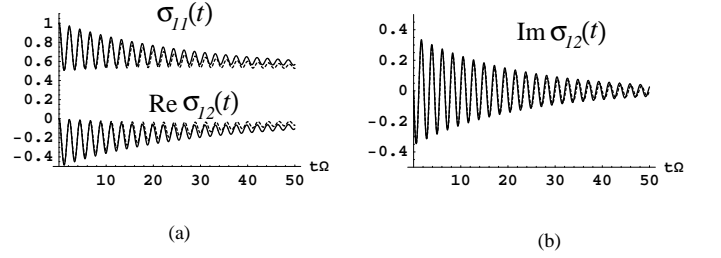}
\caption{The occupation probability of the first dot of the qubit
for $\epsilon =2\Omega$, $\Gamma_L=\Omega$,
$\Gamma_R=2\Omega$ and $\delta\Omega =0.5\Omega$. The solid line is
the exact result, whereas the dashed line is obtained from the
Bloch-type rate equations with the decoherence rate given by
Eq.~(\ref{a10}).} \label{fig4}
\end{figure}

\subsection{Fluctuation of the energy level}

Consider the SET placed near one of the qubit dots, as shown in
Fig.~2b. In this case the qubit-SET interaction term is given by
Eq.~(\ref{a1c}). As a result the energy level $E_1$ will fluctuate
under the influence of the fluctuations of the electron charge
inside the SET. The available discrete states of the entire system
are shown in Fig.~5. Using Eqs.~(\ref{a4}) we can write the rate
equations, similar to Eqs.~(\ref{a5}),
%\begin{widetext}
\begin{subequations}
\label{a11}
\begin{eqnarray}
\dot\sigma_{aa}&=&-\Gamma'_L\sigma_{aa}+\Gamma'_R\sigma_{bb}
-i\Omega_0(\sigma_{ac}-\sigma_{ca}),\\
\label{a11a}
\dot\sigma_{bb}&=&-\Gamma'_R\sigma_{bb}+\Gamma'_L\sigma_{aa}
-i\Omega_0(\sigma_{bd}-\sigma_{db}),\\
\label{a11b}
\dot\sigma_{cc}&=&-\Gamma_L\sigma_{cc}+\Gamma_R\sigma_{dd}
-i\Omega_0(\sigma_{ca}-\sigma_{ac}),\\
\label{a11c}
\dot\sigma_{dd}&=&-\Gamma_R\sigma_{dd}+\Gamma_L\sigma_{cc}
-i\Omega_0(\sigma_{db}-\sigma_{bd}),\\
\label{a11d} \dot\sigma_{ac}&=&-i\epsilon_0\sigma_{ac} -i\Omega_0
(\sigma_{aa}-\sigma_{cc})
-{\Gamma_L+\Gamma'_L\over 2}\sigma_{ac}\nonumber\\&&~~~~~~~~~~~~~~~~~~~~~~~~~~~~~~~~
+\sqrt{\Gamma_R\Gamma'_R}\sigma_{bd},\\
\label{a11e} \dot\sigma_{bd}&=&-i(\epsilon_0+U)\sigma_{bd}-
i\Omega_0(\sigma_{bb}-\sigma_{dd}) -{\Gamma_R+\Gamma'_R\over
2}\sigma_{bd}
\nonumber\\&&~~~~~~~~~~~~~~~~~~~~~~~~~~~~~~~~
+\sqrt{\Gamma_L\Gamma'_L}\sigma_{ac}\, , \label{a11f}
\end{eqnarray}
\end{subequations}
%\end{widetext}
where $\Gamma'_{L,R}$ are the tunneling rate at the
energy $E_0+U$ \cite{fn1}.

Let us assume that $\Gamma'_{L,R}=\Gamma_{L,R}$. Then it follows
from Eqs.~(\ref{a11}) that the behavior of the charge inside the SET
is not affected by the qubit, the same as in the previous case of
the Rabi frequency fluctuations. Also the qubit density matrix
becomes the mixture (\ref{a04}) in the stationary state for any
values of the qubit and the SET parameters. Hence, there is no qubit
relaxation in this case either (except for the static qubit,
$\Omega_0=0$, and $\sigma_{11}(0)\not =\sigma_{22}(0)$,
Eq.~(\ref{static})).
\begin{figure}
\includegraphics[width=9cm]{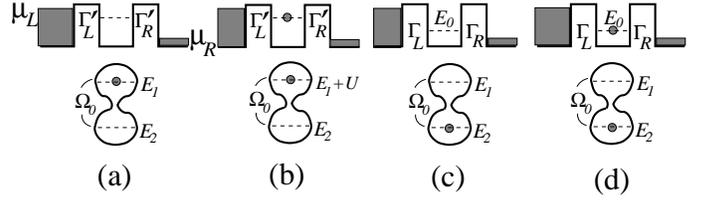}
\caption{The available discrete states of the entire system for the
configuration shown in Fig.~2b. Here $U$ is the repulsion energy
between the electrons.} \label{fig5}
\end{figure}

Since according to Eq.~(\ref{a7}), the probability of
finding an electron inside the SET in the stationary state is $\bar
P_1=\Gamma_L/\Gamma$, the energy level $E_1$ of the qubit is shifted
by $\bar P_1 U$. Therefore it is useful to define the
``renormalized'' level displacement, $\epsilon=\epsilon_0 +\bar P_1
U$.

As in the previous case we use the Laplace transform, $\sigma
(t)\to\tilde\sigma (E)$, in order to determine the decoherence rate
analytically. In the case of $\Gamma_L=\Gamma_R=\Gamma /2$ and
$\epsilon =0$ we obtain from Eqs.~(\ref{a11})
\begin{align}
\tilde\sigma_{11}(E)=\frac{i}{2E}+\frac{i}{\displaystyle 2E+\frac{32
   (E+i\Gamma ) \Omega_0 ^2}{U^2-4 E
   (E+i\Gamma )}}.
\label{a12}
\end{align}
The position of the pole in the second term of this expression
determines the decoherence rate. In contrast with Eq.~(\ref{a6}),
however, the exact analytical expression for the decoherence rate
($\Gamma_d$) is complicated, since it is given by a cubic equation.
We therefore evaluate $\Gamma_d$ in a different way, by substituting
$E=\pm 2\Omega_0-i\gamma$ in the second term of Eq.~(\ref{a12}) and
then expanding the latter in powers of $\gamma$ by keeping only the
first two terms of this expansion. The decoherence rate $\Gamma_d$
is related to $\gamma$ by $\Gamma_d=4 \gamma$, as follows from
Eq.~(\ref{aa08}). Then we obtain:
\begin{align}\label{a13}
\Gamma_d=\left\{\begin{array}{cc}{\displaystyle
U^2\Gamma\over\displaystyle 2(\Gamma^2+4\Omega_0^2)}& {\mbox{for}}
~~U\ll (\Omega_0^2+\Gamma\Omega_0)^{1/2}\\ {\displaystyle
64\Gamma\Omega_0^2\over\displaystyle U^2+16\Omega_0^2}& {\mbox{for}}
~~U\gg (\Omega_0^2+\Gamma\Omega_0)^{1/2}\end{array}\right.
\end{align}

In general, if $\Gamma_L\not =\Gamma_R$, one finds from
Eqs.~(\ref{a11}) that $\Gamma_d=2U^2\Gamma_L\Gamma_R/[\Gamma
(\Gamma^2+4\Omega_0^2)]$ for $U\ll
(\Omega_0^2+\Gamma\Omega_0)^{1/2}$. The same as in the previous case,
Eq.~(\ref{a10}), the decoherence rate
in a weak coupling limit is related to the fluctuation spectrum of
the SET, $S_Q(\omega )$, Eq.~(\ref{ac6}), but now taken at a different
frequency, $\omega =2\Omega_0$. The latter corresponds to the level
splitting of the diagonalized qubit's Hamiltonian, $\omega_R$. Thus,
\begin{align}
\Gamma_d=U^2\,S_Q(\omega_R)\, , \label{a14}
\end{align}
which can be applied also for $\epsilon\not =0$. This is illustrated
by Fig.~6 which shows $\sigma_{11}(t)$ obtained from
Eqs.~(\ref{a11}) and (\ref{aa5}) (solid line) with Eqs.~(\ref{a02a})
and (\ref{a03}) (dashed line) for the decoherence rate $\Gamma_d$
given by Eq.~(\ref{a14}). As in the previous case, shown in Fig.~4,
the initial conditions correspond to $\sigma_{11}(0)=1$ and
$\sigma_{12}(0)=0$ (respectively, $\sigma_{aa}(0)=\Gamma_R/\Gamma$
and $\sigma_{bb}(0)=\Gamma_R/\Gamma$). One finds from Fig.~6 that
Eq.~(\ref{a14}) can be used for an estimation of $\Gamma_d$ even for
$U\sim\Gamma, \Omega_0$.

In contrast with the tunneling-coupling fluctuations, Eq.~(\ref{a10}),
where the decoherence rate is given by $S_Q(0)$, the fluctuations
of the qubit's energy level generate the
decoherence rate, determined by the fluctuation spectrum at Rabi
frequency, $S_Q(\omega_R)$, Eq.~(\ref{a14}). A similar distinction
between the decoherence rates generated by different components of the
fluctuating field, exists in a phenomenological description of
magnetic resonance \cite{slich}. One can understand this distinction
by diagonalizing the qubit's Hamiltonian. In this case the Rabi
frequency, $\omega_R$, becomes the level splitting of the qubit's states
$|\pm\rangle=(|1\rangle\pm |2\rangle )/\sqrt{2}$ (for $\epsilon
=0$). So in this basis, the tunneling-coupling fluctuations correspond to
simultaneous fluctuations of the energy levels in the both dots. Since
these fluctuations are ``in phase'', we could expect that the
corresponding dephasing rate is determined by spectral density at
zero frequency. In fact, it looks like as fluctuations of a single
dot state, considered by Levinson in a weak coupling limit
\cite{levins}. On the other hand by fluctuating the energy level in one of the
dots only, one can anticipate that the corresponding dephasing rate is
determined by the fluctuation spectrum at the Rabi frequency,
$\omega_R$, Eq.~(\ref{a14}), which is a frequency of the inter-dot transitions.

Since $\omega_R$ can be
controlled by the qubit's levels displacement, $\epsilon$, the relation
(\ref{a14}) can be implied by using qubit for a measurement of
the shot-noise spectrum of the
environment\cite{schoel,agu,kouw}. For instance, it can be done by
attaching a qubit to reservoirs at different chemical potentials.
The corresponding resonant current which would flow
through the qubit in this case, can be evaluated via a simple
analytical expression \cite{g2} that includes explicitly the
decoherence rate, Eq.~(\ref{a14}). Thus by measuring this
current for different level displacement of the qubit ($\epsilon_0$),
one can extract the spectral density of the fluctuating environment
acting on the qubit\cite{agu}.
\begin{figure}
\includegraphics[width=9cm]{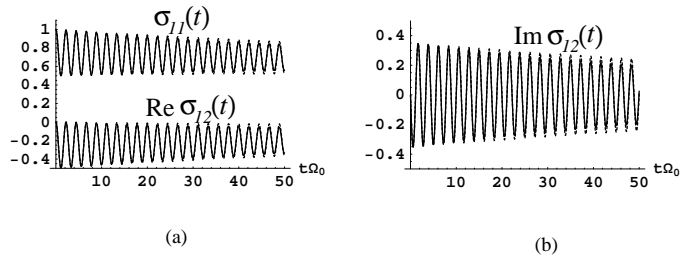}
\caption{The probability of finding the electron in the first dot of
the qubit for $\epsilon =2\Omega_0$, $\Gamma_L=\Omega_0$,
$\Gamma_R=2\Omega_0$ and $U=0.5\Omega_0$. The solid line is the
exact result, whereas the dashed line is obtained from the
Bloch-type rate equations with the decoherence rate given by
Eq.~(\ref{a14}).} \label{fig6}
\end{figure}

Although Eq.~(\ref{a14}) for the decoherence rate has been obtained
by using a particular mechanism for fluctuations of the qubit's
energy levels, we suggest that this mechanism is quite general. Indeed,
the rate equations~(\ref{a11}) can describe any fluctuating media near
a qubit, driven by the Boltzmann type of equations. Therefore
it is rather natural to assume that Eq.~(\ref{a14}) would be valid
for any type of such (classical) environment in weak coupling limit.
This implies that the decoherence rate
is always determined via the spectral density of a fluctuating
qubit's level, whereas the nature of a particular medium inducing
these fluctuations would be irrelevant. In order to substantiate
this point it is important
to compare Eq.~(\ref{a14}) with the corresponding decoherence rate
induced by the thermal environment in the framework of the spin-boson model.
In a weak damping limit this model predicts \cite{legg,weiss}
$T_1^{-1}=T_2^{-1}=(q_0^2/2)S(\omega_R)$ ,
where $q_0$ is a coupling of the medium with the qubit levels
($q_0$ corresponds to $U$ in our case) and $S(\omega )$ is a spectral density.
Using Eq.~(\ref{aa088}) one finds that this result coincides with Eq.~(\ref{a14}).

\subsection{Strong-coupling limit and localization}

Let us consider the limit of $U\gg
(\Omega_0^2+\Gamma\Omega_0)^{1/2}$. Our rate equation~(\ref{a11})
are perfectly valid in this region, providing only that $E_0+U$ is
deeply inside of the potential bias, Eq.~(\ref{aa3}). We find from Eq.~(\ref{a13})
that the
decoherence rate is not directly related to the spectrum of
fluctuations in strong coupling limit. In addition, the effective
frequency of the qubit's Rabi oscillations ($\omega_R^{eff}$)
decreases in this limit. Indeed, by using Eqs.~(\ref{a12}),
(\ref{aa7}), one finds that the main contribution to
$\sigma_{11}(t)$, is coming from a pole of $\tilde\sigma_{11}(E)$,
which lies on the imaginary axis. This implies that the effective
frequency of Rabi oscillations strongly decreases when $U\gg
(\Omega_0^2+\Gamma\Omega_0)^{1/2}$.
In addition, the decoherence rate $\Gamma_d\to 0$ in the same limit,
Eq.~(\ref{a13}). As a result, the electron would localize in the
initial qubit state, Fig.~7.

The results displayed in this figure show that the solution of the
Bloch-type rate equations, with the decoherence rate given by
Eq.~(\ref{a13}), represents damped oscillations (dashed line). It is
very far from the exact result (solid line), obtained from
Eqs.~(\ref{a11}) and corresponding to the electron localization in
the first dot. The latter is a result of an effective decrease of
the Rabi frequency for large $U$ that slows down electron
transitions between the dots. Thus such an environment-induced
localization is different from the Zeno-type effect (unlike an
assumption of Ref. \cite{hart}). Indeed, the Zeno effect takes place
whenever the decoherence rate is much larger then the coupling
between the qubit's states\cite{gfmb,g2}. However, the decoherence
rate in the strong coupling limit is much smaller then the coupling
$\Omega_0$ . In fact, the localization shown in Fig.~7 is rather
similar to that in the spin-boson model \cite{legg,weiss}. It shows
that in spite of their defferences, both models trace the same
physics of the back-action of the environment (SET) on the qubit.
\begin{figure}
\includegraphics[width=8cm]{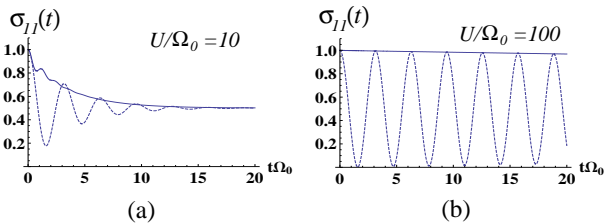}
\caption{The probability of finding the electron in the first dot of
the qubit for $\epsilon =0$, $\Gamma_L=\Gamma_R=\Omega_0$ and $U$,
as given by  Eqs.~(\ref{a11}) (solid line) and from the Bloch-type
equations (dashed line) with the decoherence rate given by
Eq.~(\ref{a13}).} \label{fig7}
\end{figure}

\section{Back-action of the qubit on the environment}
\subsection{Weak back-action effect}

Now we investigate a weak dependence of the width's $\Gamma_{L,R}$
on the energy $U$, Fig.~5. We keep only the linear term,
$\Gamma'_{L,R}=\Gamma_{L,R}+\alpha_{L,R} U$, by assuming that $U$ is
small. (A similar model has been considered in \cite{makh,shnir1}).
In contrast with the previous examples, where the widths have
not been dependent on the energy, the qubit's oscillation would
affect the SET current and its charge correlator. A more interesting
case corresponds to $\alpha_{L}\not =\alpha_R$. Let us take for
simplicity $\alpha_{L}=0$ and $\alpha_{R}=\alpha\not =0$.

Similarly to the previous case we introduce the ``renormalized''
level displacement, $\epsilon =\epsilon_0 - (\Gamma_L/\Gamma )U$,
where $\epsilon =0$ corresponds to the aligned qubit. Solving
Eqs.~(\ref{a11}) in the steady-state limit, $\bar\sigma =\sigma
(t\to\infty )$, and keeping only the first term in expansion in
powers of $U$, we find for the reduced density matrix of the qubit,
Eqs.~(\ref{aa5}):
\begin{align}
\bar\sigma =\left
(\begin{array}{cc}\displaystyle{1\over2}-{\alpha\,\epsilon\over
4\Gamma_R}&\displaystyle{\alpha\Omega_0(1+c\,\alpha\, U)\over
2\Gamma_R}
\cr\noalign{\vskip7pt}\displaystyle{\alpha\Omega_0(1+c\,\alpha\,
U)\over 2\Gamma_R} &\displaystyle{1\over2}+{\alpha\,\epsilon\over
4\Gamma_R}\end{array} \right )\, , \label{a151}
\end{align}
where $c=(\alpha\epsilon -2\Gamma )/(4\Gamma_R\Gamma)$. It follows
from Eqs.~(\ref{a151}) that the qubit's density matrix in the
steady-state is no longer a mixture, Eq.~(\ref{a04}) . Indeed, the
probability to occupy the lowest level is always larger than $1/2$
and $\bar\sigma_{12}\not =0$. This implies that relaxation takes
place together with decoherence. The ratio of the relaxation and
decoherence rates is given by the off-diagonal terms of the reduced
density matrix of the qubit. For $\epsilon =0$ one finds from
Eq.~(\ref{a09}) that $\Gamma_d/\Gamma_r=\bar\sigma^{-1}_{12}-2$.

In order to find a relation between the decoherence and relaxation
rates, $\Gamma_{d,r}$, and the fluctuation spectrum of the qubit
energy level, $S_Q(\omega )$, we first evaluate the total damping
rate of the qubit's oscillations ($\gamma$). Using Eq.~(\ref{aa08})
we find that this quantity is related to the decoherence and
relaxation rates by $\gamma=(\Gamma_d+2\Gamma_r)/4$. The same as in
the previous case the rate $\gamma$ is determined by poles of
Laplace transformed density matrix $\sigma (t)\to\tilde\sigma (E)$
in the complex $E$-plane. Consider for simplicity the case of
$\epsilon =0$ and $\Gamma_L=\Gamma_R=\Gamma /2$. Performing the
Laplace transform of Eqs.~(\ref{a11}) we look for the poles of
$\sigma_{11}(E)$ at $E=\pm 2\Omega_0-i\gamma$ by  assuming that
$\gamma$ is small. We obtain
\begin{align}
\Gamma_d+2\Gamma_r= \frac{U^2 }{2 (\Gamma ^2+4 \Omega_0^2)}\left
[\Gamma - \alpha\, U \frac{\Gamma^2-4\Omega_0^2}{2\left(\Gamma ^2+4
\Omega_0^2\right)}\right ]\label{a161}
\end{align}
for $U\ll\Omega_0$.

Now we evaluate the correlator of the charge inside the SET,
$S_Q(\omega )$ which induces the energy-level fluctuations of the
qubit. Using Eqs.~(\ref{a11}) and (\ref{ac9}) we find,
\begin{align}
S_Q(\omega )=\frac{\Gamma }{2 \left(\Gamma ^2+\omega
   ^2\right)}-\alpha\, U \frac{ \Gamma
   ^2-\omega ^2}{4 \left(\Gamma ^2+\omega
   ^2\right)^2}
\label{a171}
\end{align}
for $\alpha U\ll\Gamma$. Therefore in the limit of $U\ll\Omega_0$
and $\alpha\, U\ll\Gamma$ the total damping rate of the qubit's
oscillations is directly related to the spectral density of the
fluctuations spectrum taken at the Rabi frequency,
\begin{align}
\Gamma_d+2\Gamma_r= U^2 S_Q(2\Omega_0) .\label{a181}
\end{align}
This represents a generalization of Eq.~(\ref{a14}) for the case of
a weak back-action of qubit oscillations on the spectral density of
the environment. As a result, the qubit displays relaxation together
with decoherence. It is remarkable that the total qubit's damping
rate is still given by the fluctuation spectrum of the SET
(environment) modulated by the qubit. Note that Eq.~(\ref{a181}) can
be applied only if the modulation of the tunneling rate through the
SET (tunneling current) is small $\alpha\, U\ll\Gamma $, in addition
to a weak distortion of the qubit ($U\ll\Omega_0$).

In the case of strong back-action of the qubit on the environment
the decorerence and relaxation rates of the qubit are not directly
related to the fluctuation spectrum of the environment, even if the
distortion of the qubit is small. This point is illustrated by the
following example.

\subsection{Strong back-action}
Until now we considered the case where $E_0+U\ll \mu_L$, so that the
interacting electron of the SET remains deeply inside the voltage
bias. If however, the interaction $U$ between the qubit and the SET
is such that $E_0+U\gg\mu_L$, the qubit's oscillation would strongly
affect the fluctuation of charge inside the SET. Indeed, the current
through the SET is blocked whenever the level $E_1$ of the qubit is
occupied, Fig.~8. In fact, this case can be treated with small
modification of the rate equations (\ref{a11}), if only
$\mu_L-E_0\gg\Gamma$ and $E_0+U-\mu_L\gg\Gamma$, where $E_0$ is a
level of the SET carrying the current.

The corresponding quantum rate equations describing the system are
obtained directly from Eqs.~(\ref{a4}). Assuming that the widths
$\Gamma_{L,R}$ are energy independent we find \cite{gb}
\begin{subequations}
\label{a15}
\begin{eqnarray}
&&\dot\sigma_{aa}=(\Gamma_L+\Gamma_R)\sigma_{bb}
-i\Omega_0(\sigma_{ac}-\sigma_{ca}),\\
\label{a15a} &&\dot\sigma_{bb}=-(\Gamma_R+\Gamma_L)\sigma_{bb}
-i\Omega_0(\sigma_{bd}-\sigma_{db}),\\
\label{a15b}
&&\dot\sigma_{cc}=-\Gamma_L\sigma_{cc}+\Gamma_R\sigma_{dd}
-i\Omega_0(\sigma_{ca}-\sigma_{ac}),\\
\label{a15c}
&&\dot\sigma_{dd}=-\Gamma_R\sigma_{dd}+\Gamma_L\sigma_{cc}
-i\Omega_0(\sigma_{db}-\sigma_{bd}),\\
\label{a15d} &&\dot\sigma_{ac}=-i\epsilon_0\sigma_{ac} -i\Omega_0
(\sigma_{aa}-\sigma_{cc})
-{\Gamma_L\over 2}\sigma_{ac}\nonumber\\
&&~~~~~~~~~~~~~~~~~~~~~~~~~~~~~~~~~~~~~~~~~
+\Gamma_R\sigma_{bd},\\
\label{a15e} &&\dot\sigma_{bd}=-i(\epsilon_0+U)\sigma_{bd}-
i\Omega_0(\sigma_{bb}-\sigma_{dd})\nonumber\\
&&~~~~~~~~~~~~~~~~~~~~~~~~~~~~~
-\left(\Gamma_R+{\Gamma_L\over2}\right)\sigma_{bd}\, . \label{a15f}
\end{eqnarray}
\end{subequations}
\begin{figure}
\includegraphics[width=9cm]{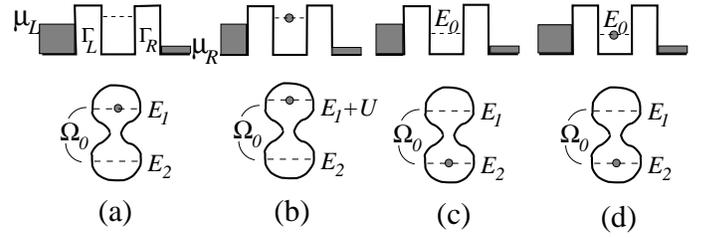}
\caption{The available discrete states of the entire system when the
electron-electron repulsive interaction $U$ breaks off the current
through the SET.} \label{fig8}
\end{figure}

Solving Eqs.~(\ref{a15}) in the stationary limit, $\bar\sigma
=\sigma (t\to\infty )$ and introducing the ``renormalized'' level
displacement, $\epsilon =\epsilon_0 - U\Gamma_L/(2\Gamma)$, we
obtain for the qubit's density matrix, Eqs.~(\ref{aa5}) in the
steady state:
\begin{subequations}
\label{a16}
\begin{eqnarray}
\bar\sigma_{11}&=&\frac{1}{2}-\frac{8 \epsilon U}{16 \epsilon^2+8 U
\epsilon+48 \Omega_0^2+9
(U^2+\Gamma^2)}\, ,\label{a16a}\\[5pt]
\bar\sigma_{12}&=&\frac{12 U \Omega _0}{16 \epsilon ^2+8 U
   \epsilon +48 \Omega _0^2+9
   \left(U^2+\Gamma ^2\right)}\, ,\label{a16b}
\end{eqnarray}
\end{subequations}
where for simplicity we considered the symmetric case,
$\Gamma_L=\Gamma_R=\Gamma /2$. It follows from Eqs.~(\ref{a16}) that
similarly to the previous example, the qubit's density matrix is no
longer a mixture (\ref{a04}). The relaxation takes place together
with decoherence in this case too.

Let us consider weak distortion of the qubit by the SET,
$U<\Omega_0$. Although the values of $U$ are restricted from below
($U\gg \Gamma +\mu_L - E_0$), this limit can be achieved if the
level $E_0$ is close to the Fermi energy, providing only that
$\mu_L-E_0\gg\Gamma$, and $\Gamma\ll U$. Now we evaluate
$\sigma_{11}(t)$ with the rate equations~(\ref{a15}) and then
compare it with the same quantity obtained from the Bloch equations,
Eq.~(\ref{aa08}), where $\Gamma_{d,r}$ are given by
Eqs.~(\ref{a14})and (\ref{a09}). The corresponding
charge-correlator, $S_Q(\omega_R)$, is evaluated by Eqs.~(\ref{ac9})
and (\ref{a15}). As an example, we take symmetric qubit with aligned
levels, $\epsilon =0$, $\Gamma_L=\Gamma_R=0.05\Omega_0$ and
$U=0.5\Omega_0$. The decoherence and relaxation rates, corresponding
to these parameters are respectively: $\Gamma_d/\Omega_0=0.0038$ and
$\Gamma_r/\Omega_0=0.00059$.

The results are presented in Fig.~9a. The solid line shows
$\sigma_{11}(t)$, obtained from the rate equations~(\ref{a15}),
where the dashed line is the same quantity obtained from
Eq.~(\ref{aa08}). We find that Eq.~(\ref{a14}) (or (\ref{a181}))
underestimates the actual damping rate of $\sigma_{11}(t)$ by an
order of magnitude). This lies in a sharp contrast with the previous
case, where the energy level of the SET is not distorted by the
qubit, $\Gamma'_{L,R}=\Gamma_{L,R}$, Fig.~5. Indeed, in this case
$\sigma_{11}(t)$ obtained Eq.~(\ref{aa08}) with $\Gamma_d$ given by
Eq.~(\ref{a14}) and $\Gamma_r=0$, agrees very well with that
obtained from the rate equations~(\ref{a11}), as shown in
Fig.~9b.
\begin{figure}
\includegraphics[width=9cm]{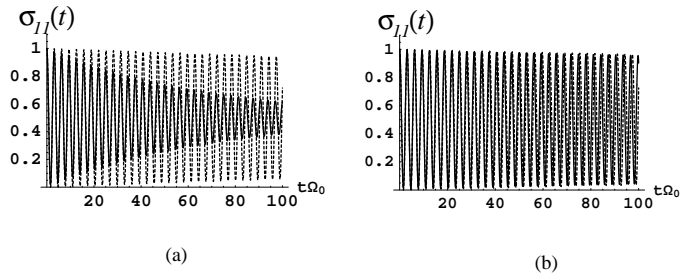}
\caption{(a) The probability of finding the electron in the first
dot of the qubit for $\epsilon =0$, $\Gamma_L=\Gamma_R=0.05\Omega_0$
and $U=0.5\Omega_0$. The solid line is obtained from
Eqs.~(\ref{a15}), whereas the dashed line corresponds to the
Eq.~(\ref{aa08}) with $\Gamma_d$ given by Eq.~(\ref{a14}); (b)  the
same for the case, shown in Fig.~5, where the solid line corresponds
to Eqs.~(\ref{a11}).} \label{fig9}
\end{figure}

Such an example clearly illustrates that the decoherence is not related
to the fluctuation spectrum of the environment, whenever the environment
is strongly affected by the qubit, even if the coupling with a qubit
is small. This is a typical case of measurement, corresponding
to a noticeable response of the environment to the qubit's state
(a ``signal'').

\section{summary}

In this paper we propose a simple model describing a qubit
interacting with fluctuating environment. The latter is represented
by a single electron transistor (SET) in close proximity of the
qubit. Then the fluctuations of the charge inside the SET generate
fluctuating field acting on the qubit. In the limit of large bias
voltage, the Schr\"odinger equation for the entire system is reduced
to the Bloch-type rate equations. The resulting equations are very
simple, so that one can easily analyze the limits of weak and strong
coupling of the qubit with the SET.

We considered separately two different cases: (a) there is no
back-action of the qubit on the SET behavior, so that the latter
represents a ``pure environment''; and (b) the SET behavior depends
on the qubit's state. In the latter case the SET can ``measure'' the
qubit. The setup corresponding to the ``pure environment'' is
realized when the energy level of the SET carrying the current lies
deeply inside the potential bias. The second (measurement) regime of
the SET is realized when the tunneling widths of the SET are energy
dependent, or when the energy level of the SET carrying the current
is close enough to the Fermi level of the corresponding reservoir.
Then the electron-electron interaction between the qubit and the SET
modulates the electron current through the SET.

In the case of the ``pure environment'' (``no-measurement'' regime)
we investigate separately two different configurations of the qubit
with respect to the SET. In the first one the SET produces
fluctuations of the off-diagonal coupling (Rabi frequency) between
two qubit's states. In the second configuration the SET produces
fluctuations of the qubit's energy levels. In the both cases we find
no relaxation of the qubit, despite the energy transfer between the
qubit and the SET can take place. As a result the qubit always turns
asymptotically to the statistical mixture. We also found that in
both cases the decoherence rate of the qubit in the weak coupling
limit is given by the spectral density of the corresponding
fluctuating parameter. The difference is that in the case of the
off-diagonal coupling fluctuations the spectral density is taken at
zero frequency, whereas in the case of the energy level fluctuations
it is taken at the Rabi-frequency.

In the case of the strong coupling limit, however, the decoherence
rate is not related to the fluctuation spectrum. Moreover we found
that the electron in the qubit is localized in this limit due to an
effective decrease of the off-diagonal coupling. This phenomenon may
resemble the localization in the spin-boson model in the strong
coupling limit.

If the charge correlator and the total SET current are affected by
the qubit (back-action effect), we found that the off-diagonal
density-matrix elements of the qubit survive in the steady-state
limit and therefore the relaxation rate is not zero. We concentrated
on the case of weak coupling, when the Coulomb repulsion between the
qubit and the SET is smaller then the Rabi frequency. The
back-action of the qubit on the SET, however, can be weak or strong.
In the first case we found that the total damping rate of the qubit
due to decoherence and relaxation is again given by the spectral
density of the SET charge fluctuations, {\em modulated by the
qubit}. This relation, however, is not working if the back-action is
strong. Indeed, we found that the damping rate of the qubit in this
case is larger by an order of magnitude than that given by the
spectral density of the corresponding fluctuating parameter.

This looks like that in the strong back-action of the qubit on the
SET the major component of decoherence is not coming from the
fluctuation spectrum of the qubit's parameters only, but also from
the measurement ``signal'' of the SET. On the first sight it could
agree with an analysis of Ref.~\cite{kor1}, suggesting that the
decoherence rate contains two components, generated by a measurement
and by a ``pure environment'' (environmental fluctuations). The
latter therefore represents an unavoidable decoherence, generated by
any environment. Yet, in a weak coupling regime such a separation
seems not working. In this case the damping (decoherence) rate is
totally determined by the environment fluctuations, even so
modulated by the qubit.

Although our model deals with a particular setup, it bears the main
physics of a fluctuating environment, acting on a qubit. Indeed, the
Bloch-type rate equations, which we used in our analysis have a
pronounced physical meaning: they relate the variation of
qubit parameters with a nearby fluctuating field described by rate
equations. A particular mechanism, generated this field should not be relevant
for an evaluations of the decoherence and relaxation rates, but only
its fluctuation spectrum. Indeed, in the weak coupling limit our result
for the decorence rate coincides with that obtained in a framework of
the spin-boson model. Thus our model can be considered as
a generic one. Its main advantage is that it can be easily extended
to multiple coupled qubits. Such an analysis would allow to
determine how decoherence scales with number of qubits \cite{zagos},
which is extremely important for a realization of quantum computations.

In addition, our model can be extended to a more complicated
fluctuating environments, such as containing characteristic
frequencies in its spectrum. It would formally correspond to a
replacement of the SET in Fig.~2 by a double-dot (DD) coupled to the
reservoirs \cite{gg}. All these situations, however, must be a
subject of a separate investigation.
\section{Acknowledgement}
One of us (S.G.) thanks T. Brandes and C. Emary for helpful
discussions and important suggestions. S.G is also grateful to the
Max Planck Institute for the Physics of Complex Systems, Dresden,
Germany, and to NTT Basic Research Laboratories, Atsugi-shi,
Kanagawa, Japan, for kind hospitality.

\appendix
\section{Quantum-mechanical derivation of rate equations for
quantum transport}

Consider the resonant tunneling through the SET, shown schematically
in Fig.~10. The entire system is described by the Hamiltonian
$H_\rset$, given by Eq.~(\ref{a1a}). The wave function can be
written in the same way as Eq.~(\ref{a2}), where the variables
related to the qubit are omitted,
\begin{align}
|\Psi (t)\rangle = \Big [ b(t) &+ \sum_l b_{0l}(t)
           c_{0}^{\dagger}c_{l}+ \sum_{l,r} b_{rl}(t)
           c_{r}^{\dagger}c_{l}\nonumber\\[5pt]
&+\sum_{l<l',r} b_{0rll'}(t)
           c_{0}^{\dagger}c_{r}^{\dagger}c_{l}c_{l'}+\cdots\Big]|\b0\rangle.
 \label{apend2}
\end{align}
Substituting $|\Psi (t)\rangle$ into the time-dependent
Schr\"odinger equation, $i\partial_t|\Psi (t)\rangle=H_\rset |\Psi
(t)\rangle$, and performing the Laplace transform,
$\tilde{b}(E)=\int_0^{\infty}\exp (iEt)\, b(t)dt$, we obtain the
following infinite set of algebraic equations for the amplitudes
$\tilde b(E)$:
\begin{subequations}
\label{ineq}
\begin{eqnarray}
&&E \tilde{b}(E) - \sum_l \Omega_{l}\tilde{b}_{0l}(E)=i
\label{ineq1}\\
&&(E + E_{l} - E_0) \tilde{b}_{0l}(E) - \Omega_{l}
      \tilde{b}(E)\nonumber\\[5pt]
      &&~~~~~~~~~~~~~~~~~~~~~~~~~~~~
      - \sum_r \Omega_{r}\tilde{b}_{lr}(E)=0
\label{ineq2}\\
&&(E + E_{l} - E_{r}) \tilde{b}_{lr}(E) -
      \Omega_{r} \tilde{b}_{0l}(E)\nonumber\\[5pt]
      &&~~~~~~~~~~~~~~~~~~~~~~~~~
      -\sum_{l'} \Omega_{l'}\tilde{b}_{0ll'r}(E)=0
\label{ineq3}\\
&&(E + E_{l} + E_{l'} - E_0 - E_{r}) \tilde{b}_{0ll'r}(E)-
\Omega_{l'} \tilde{b}_{lr}(E)\nonumber\\[5pt]
&&~~~~~~~~+ \Omega_{l} \tilde{b}_{l'r}(E)- \sum_{r'}
\Omega_{r'}\tilde{b}_{ll'rr'}(E)=0
\label{ineq4}\\[5pt]
& &~~~~~~~~~~~~~~~~~~~~\cdots\cdots\cdots\ \nonumber
\end{eqnarray}
\end{subequations}
(The r.h.s of Eq.~(\ref{ineq1}) reflects the initial condition.)
\begin{figure}
\includegraphics[width=7cm]{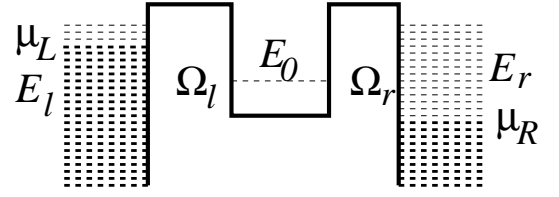}
\caption{Resonant tunneling through a single dot. $\mu_{L,R}$ are
the Fermi energies in the collector and emitter, respectively.}
\label{fig10}
\end{figure}

Let us replace the amplitude $\tilde b$ in the term
$\sum\Omega\tilde b$ of each of the equations (\ref{ineq}) by its
expression obtained from the subsequent equation. For example,
substituting $\tilde{b}_{0l}(E)$ from Eq.~(\ref{ineq2}) into
Eq.~(\ref{ineq1}) we obtain
\begin{align}
&\left [ E - \sum_l \frac{\Omega^2_{l}}{E + E_{l} - E_0}
    \right ] \tilde{b}(E)\nonumber\\[5pt]
    &~~~~~~~~~~~~~~~~~~~ - \sum_{l,r}
    \frac{\Omega_{l}\Omega_{r}}{E + E_{l} - E_0}
    \tilde{b}_{lr}(E)=i.
\label{exam}
\end{align}
Since the states in the reservoirs are very dense (continuum), one
can replace the sums over $l$ and $r$ by integrals, for instance
$\sum_{l}\;\rightarrow\;\int \rho_{L}(E_{l})\,dE_{l}\:$, where
$\rho_{L}(E_{l})$ is the density of states in the emitter, and
$\Omega_{l,r}\to\Omega_{L,R}(E_{l,r})$. Consider the first term
\begin{align}
  S_1=\int_{-\Lambda}^{\mu_L} {\Omega_L^2(E_l)\over
    E+E_l-E_0}\rho_L(E_l)dE_l
\label{exam1}
\end{align}
where $\Lambda$ is the cut-off parameter. Assuming weak energy
dependence of the couplings $\Omega_{L,R}$ and the density of states
$\rho_{L,R}$, we find in the limit of high bias,
$\mu_L=\Lambda\to\infty$
\begin{align}
  S_1=-i\pi\Omega_L^2(E_0-E)\rho_L(E_0-E)=-i{\Gamma_L\over 2}\, .
\label{exam2}
\end{align}

Consider now the second sum in Eq.~(\ref{exam}).
\begin{align}
&S_2=\int_{-\Lambda}^\Lambda\rho_R(E_r) dE_r\nonumber\\
&\times\int_{-\Lambda}^{\Lambda} {\Omega_L(E_l)\Omega_R(E_r)
    \tilde b_{lr}(E,E_l,E_r )\over E+E_l-E_0}\rho_L(E_l)
    dE_l\, ,
\label{exam3}
\end{align}
where we replaced $\tilde b_{lr}(E)$ by $\tilde b(E,E_l,E_r )$ and
took $\mu_L=\Lambda,\, \mu_R=-\Lambda$. In contrast with the first
term of Eq.~(\ref{exam}), the amplitude $\tilde b$ is not factorized
out the integral (\ref{exam3}). We refer to this type of terms as
``cross-terms''. Fortunately, all ``cross-terms'' vanish in the
limit of large bias, $\Lambda\to\infty$. This greatly simplifies the
problem and is very crucial for a transformation of the
Schr\"odinger to the rate equations. The reason is that the poles of
the integrand in the $E_l(E_r)$-variable in the ``cross-terms'' are
on the same side of the integration contour. One can find it by using a
perturbation series the amplitudes $\tilde b$ in powers of $\Omega$.
For instance, from iterations of Eqs.~(\ref{ineq}) one finds
\begin{align}
  \tilde b(E,E_l,E_r)={i\Omega_L\Omega_R
    \over E(E+E_l-E_r)(E+E_l-E_0)}+\cdots
 \label{exam4}
\end{align}

The higher order powers of $\Omega$ have the same structure. Since
$E\to E+i\epsilon$ in the Laplace transform, all poles of the
amplitude $\tilde b(E,E_l,E_r)$ in the $E_l$-variable are below the
real axis. In this case, substituting Eq.~(\ref{exam4}) into
Eq.~(\ref{exam3}) we find
\begin{widetext}
\begin{align}
  \lim_{\Lambda\to\infty}\int_{-\Lambda}^\Lambda
    \left [{\Omega_L\Omega_R
    \over (E+i\epsilon )( E+E_0-E_1+i\epsilon )^2( E+E_0-E_r+i\epsilon )}
    +\cdots\right ]dE_l=0\, ,
\label{exam31}
\end{align}
\end{widetext} Thus, $S_2\to 0$ in the limit of $\mu_L\to\infty$,
$\mu_R\to -\infty$.

Applying analogous considerations to the other equations of the
system (\ref{ineq}), we finally arrive at the following set of
equations:
\begin{subequations}
\label{fineq}
\begin{eqnarray}
&& (E + i\Gamma_L/2) \tilde{b}(E)=i
\label{fineq1}\\
&& (E + E_{l} - E_0 + i\Gamma_R/2) \tilde{b}_{0l}(E)
      \nonumber\\
&&~~~~~~~~~~~~~~~~~~~~~~~~~~~~~~~- \Omega_{l} \tilde{b}(E)=0
\label{fineq2}\\
&& (E + E_{l} - E_{r} + i\Gamma_L/2) \tilde{b}_{lr}(E)\nonumber\\
&& ~~~~~~~~~~~~~~~~~~~~~~~~~~~~~~-
      \Omega_{r} \tilde{b}_{0l}(E)=0
\label{fineq3}\\
&& (E + E_{l} + E_{l'} - E_0 - E_{r} + i\Gamma_R/2)
       \tilde{b}_{0ll'r}(E)\nonumber\\[5pt]
&&~~~~~~~~~~~~~~~ -\Omega_{l'} \tilde{b}_{lr}(E)+\Omega_{l}
\tilde{b}_{l'r}(E)=0
\label{fineq4}\\[5pt]
& &~~~~~~~~~~~~~~~~~~~~~~~~\cdots\cdots\cdots \nonumber
\end{eqnarray}
\end{subequations}

Eqs.~(\ref{fineq}) can be transformed directly to the reduced
density matrix $\sigma_{jj'}^{(n,n')}(t)$, where $j=0,1$ denote the
state of the SET with an unoccupied or occupied dot and $n$ denotes
the number of electrons which have arrived at the collector by time
$t$. In fact, as follows from our derivation, the diagonal
density-matrix elements, $j=j'$and $n=n'$, form a closed system in
the case of resonant tunneling through one level, Fig.~10. The
off-diagonal elements, $j\not =j'$, appear in the equation of motion
whenever more than one discrete level of the system carry the
transport (see Eq.~(\ref{a4}). Therefore we concentrate below on the
diagonal density-matrix elements only,
$\sigma_{00}^{(n)}(t)\equiv\sigma_{00}^{(n,n)}(t)$ and
$\sigma_{11}^{(n)}(t)\equiv\sigma_{11}^{(n,n)}(t)$. Applying the
inverse Laplace transform on finds
\begin{widetext}
\begin{subequations}
\label{invlap}
\begin{eqnarray}
\sigma_{00}^{(n)}(t)&=& \sum_{l\ldots , r\ldots}\int
\frac{dEdE'}{4\pi^2}\tilde b_{\underbrace{l\cdots}_{n}
  \underbrace {r\cdots}_{n}}(E)
\tilde b^*_{\underbrace{l\cdots}_{n}
  \underbrace {r\cdots}_{n}}(E')e^{i(E'-E)t}\label{invlapa}\\
\sigma_{11}^{(n)}(t)&=& \sum_{l\ldots , r\ldots}\int
\frac{dEdE'}{4\pi^2}\tilde b_{0{\underbrace{l\cdots}_{n+1}
  \underbrace {r\cdots}_{n}}}(E)
\tilde b^*_{0{\underbrace{l\cdots}_{n+1}
  \underbrace {r\cdots}_{n}}}(E')e^{i(E'-E)t}\label{invlapb}
\end{eqnarray}
\end{subequations}
\end{widetext}
Consider, for instance, the term $\sigma_{11}^{(0)}(t)=\sum_l
|b_{0l}(t)|^2$. Multiplying Eq.~(\ref{fineq2}) by
$\tilde{b}^*_{0l}(E')$ and then subtracting the complex conjugated
equation with the interchange $E\leftrightarrow E'$ we obtain
\begin{eqnarray}
&&\int\frac{dEdE'}{4\pi^2}\sum_l\left[(E'-E -i\Gamma_R)\tilde
b_{0l}(E)\tilde b^*_{0l}(E')\right.\nonumber\\&&\left.~~-2{\mbox
{Im}}\sum_l\Omega_l \tilde b_{0l}(E)\tilde
b^*(E')\right]e^{i(E'-E)t}=0 \label{ap3}
\end{eqnarray}

Using Eq.~(\ref{invlapb}) one easily finds that the first integral
in Eq.~(\ref{ap3}) equals to $-i[\dot\sigma_{11}^{(0)}(t)+
\Gamma_R\sigma_{11}^{(0)}(t)]$. Next, substituting
\begin{align}
\tilde{b}_{0l}(E)=\frac{\Omega_{l} \tilde{b}(E)} {E + E_{l} -
E_0 + i\Gamma_R/2} \label{ap2}
\end{align}
from Eq.~(\ref{fineq2}) into the second term of Eq.~(\ref{ap3}), and
replacing a sum by an integral, one can perform the
$E_l$-integration in the large bias limit, $\mu_L\to\infty$,
$\mu_R\to -\infty$. Then using again Eq.~(\ref{invlapb}) one reduces
the second term of Eq.~(\ref{ap3}) to
$i\Gamma_L\sigma_{00}^{(0)}(t)$. Finally, Eq.~(\ref{ap3}) reads
$\dot{\sigma}^{(0)}_{11}(t)=\Gamma_L \sigma^{(0)}_{00}(t) - \Gamma_R
\sigma_{11}^{(0)}(t)$.

The same algebra can be applied for all other amplitudes $\tilde
b_\alpha (t)$. For instance, by using Eq.~(\ref{invlapa}) one easily finds
that Eq.~(\ref{fineq3}) is converted to the following rate equation
$\dot{\sigma}^{(1)}_{00}(t)=-\Gamma_L \sigma^{(1)}_{00}(t) +\Gamma_R
\sigma_{11}^{(0)}(t)$. With respect to the states involving more
than one electron (hole) in the reservoirs (the amplitudes like
$\tilde b_{0ll'r}(E)$ and so on), the corresponding equations
contain the Pauli exchange terms. By converting these equations into
those for the density matrix using our procedure, one finds the
``cross terms'', like $\sum\Omega_{l} \tilde{b}_{l'r}(E)\Omega_{l'}
\tilde{b}^*_{lr}(E')$, generated by Eq.~(\ref{fineq4}). Yet, these
terms vanish after an integration over $E_{l(r)}$ in the large bias
limit, as the second term in Eq.~(\ref{exam}). The rest of the
algebra remains the same, as described above. Finally we arrive at
the following infinite system of the chain equations for the
diagonal elements, $\sigma_{00}^{(n)}$ and $\sigma_{11}^{(n)}$, of
the density matrix,
\begin{subequations}
\label{aandb}
\begin{eqnarray}
& &\dot{\sigma}^{(0)}_{00}(t) = - \Gamma_L \sigma^{(0)}_{00}(t)\;,
\label{anought}\\
& &\dot{\sigma}^{(0)}_{11}(t) = \Gamma_L \sigma^{(0)}_{00}(t)
                               - \Gamma_R \sigma_{11}^{(0)}(t)\;,
\label{bnought}\\
& &\dot{\sigma}^{(1)}_{00}(t) = - \Gamma_L \sigma^{(1)}_{00}(t)
                               + \Gamma_R \sigma_{11}^{(0)}(t)\;,
\label{aone}\\
& &\dot{\sigma}^{(1)}_{11}(t) = \Gamma_L \sigma^{(1)}_{00}(t)
                               - \Gamma_R \sigma_{11}^{(1)}(t)\;,
\label{bone}\\
& &~~~~~~~~~~~~~~~~~\cdots\cdots\cdots \nonumber
\end{eqnarray}
\end{subequations}

Summing over $n$ in Eqs.~(\ref{aandb}) we find for the reduced density
matrix of the SET, $\sigma (t)=\sum_n\sigma^{(n)}(t)$, the following
``classical'' rate equations,
\begin{subequations}
\label{ap4}
\begin{eqnarray}
\dot\sigma_{00}(t)&=&-\Gamma_L\sigma_{00}(t)+\Gamma_R\sigma_{11}(t)
\label{ap4a}\\
\dot\sigma_{11}(t)&=&\Gamma_L\sigma_{00}(t)-\Gamma_R\sigma_{11}(t)
\label{ap4b}
\end{eqnarray}
\end{subequations}
These equations represent a particular case of our general quantum
rate equations (\ref{a4}), which are derived using the above
described technique\cite{gp,g1}.
\section{Correlator of electric charge inside the SET.}

The charge correlator inside the SET is given by $S_Q(\omega )=\bar
S_Q(\omega )+\bar S_Q(-\omega )$, where
\begin{align}
\bar S_Q(\omega )=\int_0^\infty \langle\delta \hat Q(0)\delta \hat
Q(t)\rangle e^{i\omega t}dt\, . \label{ac1}
\end{align}
Here $\delta \hat Q(t)=c_0^\dagger (t)c_0(t)-\bar q$ and $\bar
q=\bar P_1=P_1(t\to\infty )$ is the average charge inside the dot.
Since the initial state, $t=0$ in Eq.~(\ref{ac1}) corresponds to the
steady state, one can represent the time-correlator as
\begin{align}
\langle\delta \hat Q(0)\delta \hat
Q(t)\rangle=\sum_{q=0,1}P_q(0)(q-\bar q)(\langle Q_q(t)\rangle -\bar
q)\, , \label{ac2}
\end{align}
where $P_q(0)$ is the probability of finding the charge $q=0,1$
inside the quantum dot in the steady state, such that $P_1(0)=\bar
q$ and $P_0(0)=1-\bar q$, and $\langle Q_q(t)\rangle =P_1^{(q)}(t)$
is the average charge in the dot at time $t$, starting with the
initial condition $P^{(q)}_1(0) =q$. Substituting Eq.~(\ref{ac2})
into Eq.~(\ref{ac1}) we finally obtain
\begin{align}
\bar S_Q(\omega )=\bar q(1-\bar q)[\tilde P_1^{(1)}(\omega )-\tilde
P_1^{(0)}(\omega ) \label{ac3}]\, ,
\end{align}
where $\tilde P_1^{(q)}(\omega )$ is a Laplace transform of
$P_1^{(q)}(t)$. These quantities are obtained directly from the rate
equations, such that $\bar q=\bar\sigma_{bb}+\bar\sigma_{dd}$ and
$\tilde P_1^{(q)}(\omega )=\tilde\sigma^{(q)}_{bb}(\omega
)+\tilde\sigma^{(q)}_{dd}(\omega )$, where $\bar\sigma =\sigma
(t\to\infty )$ and $\tilde\sigma^{(q)}(\omega )$ is the Laplace
transform $\sigma^{(q)}(t)$ with the initial conditions
corresponding to the occupied ($q=1$) or unoccupied ($q=0$) SET. In
order to find these quantities it is useful to rewrite the rate
equations in the matrix form, $\dot\sigma (t)=M \sigma (t)$,
representing $\sigma (t)$ as the eight-vector, $\sigma
=\{\sigma_{aa},\sigma_{bb},\sigma_{cc},\sigma_{dd},\sigma_{ac},
\sigma_{ca},\sigma_{bd},\sigma_{db}\}$ and $M$ as the corresponding
$8\times 8$-matrix. Applying the Laplace transform we find the
following matrix equation,
\begin{align}
(i\,\omega\, I +M)\tilde\sigma^{(q)}(\omega )=-\sigma^{(q)}(0)\,
,\label{ac7}
\end{align}
where $I$ is the unit matrix and $\sigma^{(q)}(0)$ is the initial
condition for the density-matrix obtained by projecting the total
wave function (\ref{a2}) on occupied ($q=1$) and unoccupied ($q=0$)
states of the SET in the limit of $t\to\infty$,
\begin{subequations}
\label{ac8}
\begin{eqnarray}
\sigma^{(1)}(0)&=&{\cal
N}_1\{0,\bar\sigma_{bb},0,\bar\sigma_{dd},0,0,\bar\sigma_{bd},
\bar\sigma_{db}\}\, , \label{ac8a}\\
\sigma^{(0)}(0)&=&{\cal
N}_0\{\bar\sigma_{aa},0,\bar\sigma_{cc},0,\bar\sigma_{ac},
\bar\sigma_{ca},0,0\}\, , \label{ac8b}
\end{eqnarray}
\end{subequations}
and ${\cal N}_1=1/\bar q$ and ${\cal N}_0=1/(1-\bar q)$ are the
corresponding normalization factors. Finally one obtains:
\begin{align}
S_Q(\omega )=2\bar q(1-\bar
q){\mbox{Re}}\,[\tilde\sigma_{bb}^{(1)}&(\omega
)+\tilde\sigma_{dd}^{(1)}(\omega )\nonumber\\[5pt]
&-\tilde\sigma_{bb}^{(0)}(\omega
)-\tilde\sigma_{dd}^{(0)}(\omega )]. \label{ac9}
\end{align}

In the case shown in Fig.~2 one finds from Eqs.~(\ref{a5}) or
Eqs.~(\ref{a11}) for $\Gamma'_{L,R}=\Gamma_{L,R}$ that
$\bar\sigma_{ac}=\sigma_{bd}=0$, $\bar q=\Gamma_L/\Gamma$ and
$\tilde\sigma_{bb}^{(q)}(\omega )+\tilde\sigma_{dd}^{(q)}(\omega
)=\tilde P_1^{(q)}(\omega )$. The latter equation is given by
\begin{align}
(i\omega -\Gamma )\tilde P_1^{(q)}(\omega
)=-q+{i\Gamma_L\over\omega}\, . \label{ac4}
\end{align}
Substituting Eq.~(\ref{ac4}) into Eq.~(\ref{ac3}) one obtains:
\begin{align}
S_Q(\omega )={2\Gamma_L\Gamma_R\over\Gamma(\omega^2 +\Gamma^2)}\, .
\label{ac6}
\end{align}

Obviously, for a more general case when $\Gamma'_{L,R}\not
=\Gamma_{L,R}$, or when the electron-electron interaction excites
the electron inside the SET above the Fermi level, Fig.~8, the
expressions for $S_Q(\omega )$, obtained from Eq.~(\ref{ac9}) have a
more complicated than Eq.~(\ref{ac6}).


\begin{thebibliography}{999}
\bibitem{fn1} In a strict sense the quantum rate equations~(\ref{a4}) were
derived by assuming constant widths $\Gamma$. Yet these equation are
also valid when the widths are weakly energy dependent, as follows
from their derivations (see \cite{gp,g1} and Appendix A).
\end{thebibliography}
\end{document}